%
\let\includefigures=\iftrue
%
%
\let\useblackboard=\iftrue
%
%
\newfam\black
\input harvmac.tex
\includefigures
\message{If you do not have epsf.tex (to include figures),}
\message{change the option at the top of the tex file.}
\input epsf
\def\figin{\epsfcheck\figin}\def\figins{\epsfcheck\figins}
\def\epsfcheck{\ifx\epsfbox\UnDeFiNeD
\message{(NO epsf.tex, FIGURES WILL BE IGNORED)}
\gdef\figin##1{\vskip2in}\gdef\figins##1{\hskip.5in}
\else\message{(FIGURES WILL BE INCLUDED)}%
\gdef\figin##1{##1}\gdef\figins##1{##1}\fi}
\def\DefWarn#1{}
\def\figinsert{\goodbreak\midinsert}
\def\ifig#1#2#3{\DefWarn#1\xdef#1{fig.~\the\figno}
\writedef{#1\leftbracket fig.\noexpand~\the\figno}%
\figinsert\figin{\centerline{#3}}\medskip\centerline{\vbox{\baselineskip12pt
\advance\hsize by -1truein\noindent\footnotefont{\bf Fig.~\the\figno:} #2}}
\bigskip\endinsert\global\advance\figno by1}
\else
\def\ifig#1#2#3{\xdef#1{fig.~\the\figno}
\writedef{#1\leftbracket fig.\noexpand~\the\figno}%
\global\advance\figno by1}
\fi
\useblackboard
\message{If you do not have msbm (blackboard bold) fonts,}
\message{change the option at the top of the tex file.}
\font\blackboard=msbm10 scaled \magstep1
\font\blackboards=msbm7
\font\blackboardss=msbm5
\textfont\black=\blackboard
\scriptfont\black=\blackboards
\scriptscriptfont\black=\blackboardss

\else

\fi
%
\def\yboxit#1#2{\vbox{\hrule height #1 \hbox{\vrule width #1
\vbox{#2}\vrule width #1 }\hrule height #1 }}
\def\fillbox#1{\hbox to #1{\vbox to #1{\vfil}\hfil}}
\def\ybox{{\lower 1.3pt \yboxit{0.4pt}{\fillbox{8pt}}\hskip-0.2pt}}

\def\comments#1{}

\def\p{\partial}

\def\eps{\epsilon}
\def\half{{1\over 2}}

\def\tr{{\rm tr\ }}

\def\CN{{\cal N}}

\def\nl{\hfill\break}

\def\ap{\alpha'}
\def\sqap{\sqrt{\alpha'}}

\def\I{I}

\def\II{\relax{I\kern-.10em I}}
\def\IIa{{\II}a}
\def\IIb{{\II}b}

\def\IZ{\relax\ifmmode\mathchoice
{\hbox{\cmss Z\kern-.4em Z}}{\hbox{\cmss Z\kern-.4em Z}}
{\lower.9pt\hbox{\cmsss Z\kern-.4em Z}}
{\lower1.2pt\hbox{\cmsss Z\kern-.4em Z}}\else{\cmss Z\kern-.4em
Z}\fi}
\def\IB{\relax{\rm I\kern-.18em B}}
\def\IC{{\relax\hbox{$\inbar\kern-.3em{\rm C}$}}}
\def\ID{\relax{\rm I\kern-.18em D}}
\def\IE{\relax{\rm I\kern-.18em E}}
\def\IF{\relax{\rm I\kern-.18em F}}
\def\IG{\relax\hbox{$\inbar\kern-.3em{\rm G}$}}
\def\IGa{\relax\hbox{${\rm I}\kern-.18em\Gamma$}}
\def\IH{\relax{\rm I\kern-.18em H}}
\def\II{\relax{\rm I\kern-.18em I}}
\def\IK{\relax{\rm I\kern-.18em K}}
\def\IP{\relax{\rm I\kern-.18em P}}

%
\def\mod{{\rm mod}}

\def\p{\partial}

\font\cmss=cmss10 \font\cmsss=cmss10 at 7pt
\def\IR{\relax{\rm I\kern-.18em R}}

\def\inbar{\,\vrule height1.5ex width.4pt depth0pt}

\def\BR{\IR}
\def\BZ{\IZ}
\def\BR{\IR}
\def\BC{\IC}

\def\ls{l_s}
\def\ms{m_s}
\def\gs{g_s}
\def\lp10{l_P^{10}}
\def\lp11{l_P^{11}}
\def\R11{R_{11}}

\Title{\vbox{\baselineskip12pt\hbox{hep-th/9608024}
\hbox{RU-96-62}}}
{\vbox{
\centerline{D-branes and Short Distances} 
\smallskip
\centerline{in String Theory}}}
\centerline{Michael R. Douglas, Daniel Kabat, Philippe Pouliot, and
Stephen H. Shenker}
\medskip
\centerline{\it Department of Physics and Astronomy}
\centerline{\it Rutgers University }
\centerline{\it Piscataway, NJ 08855--0849}
\medskip
\centerline{\tt mrd, kabat, pouliot, shenker@physics.rutgers.edu}
\medskip
\bigskip
\noindent
We study the behavior of D-branes at distances far shorter than the
string length scale~$l_s$.  We argue that short-distance phenomena are
described by the IR behavior of the D-brane world-volume quantum
theory.  This description is valid until the brane motion becomes
relativistic.  At weak string coupling $\gs$ this corresponds to
momenta and energies far above string scale.  We use 0-brane quantum
mechanics to study 0-brane collisions and find structure at length
scales corresponding to the eleven-dimensional Planck length ($\lp11
\sim \gs^{1/3} l_s$) and to the radius of the eleventh dimension in
M-theory ($\R11 \sim \gs l_s$).  We use 0-branes to probe non-trivial
geometries and topologies at sub-stringy scales.  We study the 0-brane
4-brane system, calculating the 0-brane moduli space metric, and find
the bound state at threshold, which has characteristic size $\lp11$.
We examine the blowup of an orbifold and are able to resolve the
resulting $S^2$ down to size $\lp11$.  A 0-brane with momentum approaching
$1/\R11$ is able to explore a larger configuration space in which the
blowup is embedded.  Analogous phenomena occur for small instantons.
We finally turn to 1-branes and calculate the size of a bound state to
be $\sim \gs^{1/2} l_s$, the 1-brane tension scale.
\Date{August 1996}
\def\half{{1 \over 2}}
\def\identity{{\rlap{\cmss 1} \hskip 1.6pt \hbox{\cmss 1}}}
\def\laplace{{\kern1pt\vbox{\hrule height 1.2pt\hbox{\vrule width 1.2pt\hskip
  3pt\vbox{\vskip 6pt}\hskip 3pt\vrule width 0.6pt}\hrule height 0.6pt}
  \kern1pt}}
\def\scriptlap{{\kern1pt\vbox{\hrule height 0.8pt\hbox{\vrule width 0.8pt
  \hskip2pt\vbox{\vskip 4pt}\hskip 2pt\vrule width 0.4pt}\hrule height 0.4pt}
  \kern1pt}}

\def\integer{{\rlap{\cmss Z} \hskip 1.6pt \hbox{\cmss Z}}}

\def\np{{\it Nucl.~Phys.~}}
\def\pl{{\it Phys.~Lett.~}}

\def\cmp{{\it Commun.~Math.~Phys.~}}

\def\mpl{{\it Mod.~Phys.~Lett.~}}

\lref\shenker{S.~H.~Shenker, {\it Another Length Scale in String Theory?},
hep-th/9509132.}
\lref\dlp{J.~Dai, R.~G.~Leigh and J.~Polchinski, {\it Mod.~Phys.~Lett.}
{\bf A4} (1989) 2073.}
\lref\pol{J.~Polchinski, {\it Phys.~Rev.~Lett.~}{\bf 75} (1995) 4724,
hep-th/9510017.}
\lref\joerev{S.~Chaudhuri, C.~Johnson, and J.~Polchinski,
{\it Notes on D-Branes}, hep-th/9602052.}
\lref\kp{D.~Kabat and P.~Pouliot,
{\it A Comment on Zero-Brane Quantum Mechanics}, hep-th/9603127.}
\lref\KleTho{I.~R.~Klebanov and L.~Thorlacius, {\it The Size of p-Branes},
hep-th/9510200.}
\lref\CalKle{C.~G.~Callan, Jr.~and I.~R.~Klebanov,
{\it D-Brane Boundary State Dynamics}, hep-th/9511173.}
\lref\KhuMye{R.~R.~Khuri and R.~C.~Myers, {\it Low-Energy Scattering
of Black Holes and $p$-branes in String Theory}, hep-th/9512137.}
\lref\Gub{S.~S.~Gubser, A.~Hashimoto, I.~R.~Klebanov, and J.~M.~Maldacena,
{\it Gravitational lensing by $p$-branes}, hep-th/9601057.}
\lref\HKleb{A.~Hashimoto and I.~R.~Klebanov, {\it Decay of Excited D-branes},
hep-th/9604065.}
\lref\Barbon{J.~L.~F. Barb\'on, {\it D-brane
Form-Factors at High-Energy}, hep-th/9601098.}
\lref\GibMan{G.~W.~Gibbons and N.~S.~Manton, {\it Nucl.~Phys.~} {\bf B274}
(1986) 183.}
\lref\Senbdmon{A.~Sen, {\it Phys.~Lett.} {\bf B329} (1994) 217.}
\lref\dl{M.~R.~Douglas and M.~Li, {\it D-brane Realization of N=2
Super Yang-Mills Theory in Four Dimensions}, hep-th/9604041.}
\lref\koblitz{N.~Koblitz, {\it Introduction to Elliptic Curves and Modular
Forms} (Springer-Verlag, 1994).}
\lref\dkaz{M.~R.~Douglas and V.~A.~Kazakov, {\it Phys.~Lett.~} {\bf B319}
(1993) 219.}
\lref\witten{E.~Witten, {\it Small Instantons in String Theory},
hep-th/9511030.}
\lref\witadhm{E.~Witten, {\it Sigma Models and the ADHM Construction of
Instantons}, {\it J.~Geom.~Phys.~} {\bf 15} (1995) 215.}
\lref\gimon{E.~G.~Gimon and J.~Polchinski,
{\it Consistency Conditions for Orientifolds and D-manifolds}, hep-th/9601038.}
\lref\douglas{M.~R.~Douglas, {\it Branes within Branes}, hep-th/9512077.}
\lref\dm{M.~R.~Douglas and G.~Moore, {\it D-Branes, Quivers, and ALE
Instantons},
hep-th/9603167.}
\lref\dgauge{M.~R.~Douglas, {\it Gauge Fields and D-Branes},  hep-th/9604198.}
\lref\polwit{J.~Polchinski and E.~Witten, {\it Evidence for Heterotic-Type I
String Duality},  hep-th/9510169.}
\lref\zwiebach{B.~Zwiebach, {\it Phys.~Lett.~} {\bf B256} (1991) 22.}
\lref\WitSFT{E.~Witten, \np {\bf B268} (1986) 253.}
\lref\compsource{J.~Polchinski and Y.~Cai, {\it Nucl.~Phys.~}
{\bf B296} (1988) 91;\nl C.~G.~Callan, C.~Lovelace, C.~R.~Nappi and S.~A.~Yost,
{\it Nucl.~Phys.} {\bf B308} (1988) 221;\nl
M.~Li, {\it Boundary States of D-Branes and Dy-Strings},
hep-th/9510161.}
\lref\ghm{M.~Green, J.~Harvey and G.~Moore,
{\it I-Brane Inflow and Anomalous Couplings on D-Branes},  hep-th/9605033.}
\lref\hs{G.~T.~Horowitz and A.~Strominger, {\it Nucl.~Phys.~} {\bf B360}
(1991) 197.}
\lref\VafaF{C.~Vafa, {\it Evidence for F-Theory}, hep-th/9602022.}
\lref\SenF{A.~Sen, {\it F-theory and Orientifolds}, hep-th/9605150.}
\lref\SenBound{A.~Sen, {\it A Note on Marginally Stable Bound States in Type
II String Theory}, hep-th/9510229.}
\lref\BanDS{T.~Banks, M.~R.~Douglas, and N.~Seiberg,
{\it Probing F-theory with Branes}, hep-th/9605199.}
\lref\SeiI{N.~Seiberg, {\it IR Dynamics on Branes and Space-Time Geometry},
hep-th/9606017.}
\lref\DeWitLuNi{B.~de Wit, M.~L\"uscher, and H.~Nicolai, \np {\bf B320}
(1989) 135.}
\lref\ShenkerNP{S.~H.~Shenker, in {\it Random Surfaces and Quantum Gravity},
eds.~O.~Alvarez, E.~Marinari, and P.~Windey (Plenum, 1991).}
\lref\polrev{For a review of the recent dramatic progress in
our understanding of string theory, see  J.~Polchinski, {\it String
Duality--A Colloquium}, hep-th/9607050.}
\lref\Schwarz{J.~Schwarz, {\it An $SL(2,\integer)$ Multiplet of Type IIB
Superstrings}, hep-th/9508143.}
\lref\WittenBound{E.~Witten, {\it Bound States of Strings and p-Branes},
hep-th/9510135.}
\lref\Schmid{C.~Schmidhuber, {\it D-brane actions}, hep-th/9601003.}
\lref\Bachas{C.~Bachas, {\it D-Brane Dynamics}, hep-th/9511043.}
\lref\Lifschytz{G.~Lifschytz, {\it Comparing D-branes to Black-branes},
hep-th/9604156.}
\lref\Sakurai{J.~J.~Sakurai, {\it Advanced Quantum Mechanics} (Addison-Wesley,
1967), p.~119.}
\lref\AndTsey{O.~D.~Andreev and A.~A.~Tseytlin, \np {\bf B311} (1988) 205.}
\lref\WittenGrass{E.~Witten, {\it The Verlinde Algebra and the Cohomology
of the Grassmannian}, hep-th/9312104, p.~71.}
\lref\DaiLeiPol{J.~Dai, R.~G.~Leigh, and J.~Polchinski, \mpl
{\bf A4} (1989) 2073.}
\lref\Lei{R.~G.~Leigh, \mpl {\bf A4} (1989) 2767.}
\lref\PolBdy{J.~Polchinski, {\it Combinatorics of Boundaries in String
Theory}, hep-th/9407031.}
\lref\PolRR{J.~Polchinski, {\it Dirichlet-Branes and Ramond-Ramond
Charges}, hep-th/9510017.}
\lref\Green{M.~B.~Green, {\it Phys.~Lett.~} {\bf B69} (1977) 89;
{\bf B201} (1988) 42; {\bf B282} (1992) 380; {\bf B329} (1994) 435.}
\lref\Li{M.~Li, {\it Boundary States of D-Branes and Dy-Strings},
hep-th/9510161\semi
{\it Dirichlet Boundary State in Linear Dilaton Background},
hep-th/9512042.}
\lref\KleTho{I.~R.~Klebanov and L.~Thorlacius, {\it The Size of p-Branes},
hep-th/9510200.}
\lref\CalKle{C.~G.~Callan, Jr.~and I.~R.~Klebanov,
{\it D-Brane Boundary State Dynamics}, hep-th/9511173.}
\lref\BanSus{T.~Banks and L.~Susskind, {\it Brane-Antibrane Forces},
hep-th/9511194.}
\lref\BSVDman{M.~Bershadsky, C.~Vafa, and V.~Sadov,
{\it D-Strings on D-Manifolds}, hep-th/9510225.}
\lref\BerVafSad{M.~Bershadsky, C.~Vafa, and V.~Sadov, {\it D-Branes
and Topological Field Theories}, hep-th/9511222.}
\lref\TownI{P.~K.~Townsend, {\it D-branes from M-branes}, hep-th/9512062.}
\lref\KapMich{D.~Kaplan and J.~Michelson, {\it Zero Modes for the $D=11$
Membrane and Five-brane}, hep-th/9510053.}
\lref\Doug{M.~R.~Douglas, {\it Branes within Branes},
hep-th/9512077.}
\lref\Gubs{S.~S.~Gubser, A.~Hashimoto, I.~R.~Klebanov, and J.~M.~Maldacena,
{\it Gravitational lensing by $p$-branes}, hep-th/9601057.}
\lref\TownII{P.~K.~Townsend, \pl {\bf B350} (1995) 184, hep-th/9501068.}
\lref\WittenII{E.~Witten, {\it String Theory Dynamics in Various
Dimensions}, hep-th/9503124.}
\lref\FraTse{E.~S.~Fradkin and A.~A.~Tseytlin, \pl {\bf B163} (1985)
123.}
\lref\Abou{A.~Abouelsaood, C.~G.~Callan, C.~R.~Nappi, and S.~A.~Yost,
\np {\bf B280} (1987) 599.}
\lref\KutSei{D.~Kutasov and N.~Seiberg, {\it Nucl.~Phys.~}
{\bf B358} (1991) 600.}
\lref\Tse{A.~A.~Tseytlin, {\it Self-Duality of Born-Infeld action and
Dirichlet 3-brane of type IIB superstring theory}, hep-th/9602064.}
\lref\DanFerSun{U.~H.~Danielsson, G.~Ferretti, and B.~Sundborg, {\it
D-particle Dynamics and Bound States}, hep-th/9603081.}
\lref\PolWit{J.~Polchinski and E.~Witten, {\it Evidence for Heterotic-Type
I String Duality}, hep-th/9510169.}
\lref\eightb{E.~Bergshoeff, M.~de~Roo, M.~B.~Green, G.~Papadopoulos,
and P.~K.~Townsend, {\it Duality of Type II 7-branes and 8-branes,}
hep-th/9601150.}
\lref\PolNew{J.~Polchinski, {\it Tensors from K3 Orientifolds},
hep-th/9606165.}
\lref\AGM{P.~Aspinwall, B.~Greene and D.~R.~Morrison, \np {\bf B416} (1994)
414, hep-th/9309097; \np {\bf B420} (1994) 184, hep-th/9311042.}
\lref\Aspinwall{P.~Aspinwall, {\it Resolution of Orbifold Singularities in
String Theory}, to appear in {\it Essays on Mirror Manifolds 2},
hep-th/9403123.}
\lref\WitPhase{E.~Witten, {\it Phases Of N=2 Theories In Two Dimensions},
\np {\bf  B403} (1993) 159, hep-th/9301042.}
\lref\Strom{A.~Strominger, \np {\bf B451} (1995) 96, hep-th/9504090 \semi
B.~Greene, D.~R.~Morrison and A.~Strominger,
\np {\bf B451} (1995) 109, hep-th/9504145.}
\lref\Kronheimer{P.~Kronheimer, {\it J.~Diff.~Geom.~} {\bf 28} (1989) 665.}
\lref\HKLM{A.~Hitchin, A.~Karlhede, U.~Lindstrom and M.~Ro{\v c}ek,\nl
\cmp {\bf 108} (1987) 535.}
\lref\BlackHoles{A.~Strominger and C.~Vafa, {\it Microscopic
Origin of the Bekenstein-Hawking Entropy}, hep-th/9601029.}
\lref\friedan{D.~Friedan, {\it Ann. Phys.} {\bf 163} (1985) 318.}
\lref\toappear{T.~Banks, M.~R.~Douglas, J.~Polchinski, S.~Shenker and
A.~Strominger, work in progress.}
\lref\Superspace{S.~J.~Gates, Jr., M.~T.~Grisaru, M.~Ro{\v c}ek,
and W.~Siegel, {\it Superspace} (Benjamin-Cummings, 1983), pp. 454-455.}
\lref\BergRooOrtin{E.~Bergshoeff, M.~de Roo, and T.~Ort\'\i n, {\it The
Eleven-Dimensional Five-Brane}, hep-th/9606118.}
\lref\Aspinwallmin{P. Aspinwall, {\it Nucl. Phys.} {\bf B431} (1994) 78. }
\lref\birdav{N.~D.~Birrell and P.~C.~W. ~Davies, {\it Quantum Fields
in Curved Space} (Cambridge, 1982), p. 228.}
\lref\SethiStern{S.~Sethi and M.~Stern, {\it A Comment
on the Spectrum of H-Monopoles}, hep-th/9607145.}
\lref\GauntHarv{J.~Gauntlett and J.~Harvey, {\it 
S-Duality and the Spectrum of Magnetic
Monopoles in Heterotic String Theory}, hep-th/9407111.}

\newsec{Introduction}

Recent developments \polrev\ have brought an interesting new regime of
string theory under control: processes involving the non-relativistic
dynamics of Dirichlet branes.  As we shall see, such processes can be
used to directly probe the short-distance behavior of string
theory.

It has often been proposed that in string theory our conventional
ideas of space-time cease to make sense below a minimum length: the
string scale, $l_s \sim \sqap \sim 1/m_s $.  We will present
evidence that space-time does make sense below the string scale.
Furthermore, we will discuss specific modifications to the
conventional ideas of geometry, which are appropriate for the
processes we consider.

In quantum gravity, the Planck length (at weak coupling far smaller
than the string length) is often posited as a minimum length.  In
particular, the black hole entropy formula suggests that the Planck
length sets a limit on the density of information.  The results we
discuss here imply the existence of scales far shorter than the 
ten-dimensional Planck length.  The way in which our results, which apply
to a small number of D-branes, can be reconciled with semiclassical
black hole computations is an important question for future
investigation.

In string theory, most previous results which found evidence that the
minimum length is the string scale used strings themselves as probes.
Until recently it was difficult to do otherwise.  The intuitive
picture is that increasing the momentum, which allows probing smaller
distance scales, requires increasing the energy.  But this energy is
converted into string tension, increasing the size of the probe.  More
generally, if no object in the theory has size smaller than $l_s$, it
is hard to make any simple proposal for what smaller scale structure
would mean.

In the light of string duality, solitons are equally natural probes.
Some of these are large classical solutions which would seem even less
suited for studying these questions.  However in \shenker\ evidence
was presented that solitons with RR charge are much smaller than the
string scale when the string coupling $g_s$ is small.  A heuristic
indication of this follows from the charge quantization condition for
RR solitons, which is $\int F = n g_s$ where $F$ is a RR field
strength in string normalization.  At weak coupling the charge of such
a soliton is anomalously small, compared to a conventional NS soliton.
Since they are BPS saturated, their tension ($\sim 1/g_s$ \WittenII),
though large in the weak coupling limit (allowing them to be
localized), is also anomalously small compared to an NS soliton.  Thus
their gravitational, gauge, and other fields become strong only at
sub-stringy distances.  This small size makes RR solitons natural
probes of sub-stringy structure.

The key realization that RR solitons are represented by D-branes makes
them very accessible to world-sheet computation
\refs{\pol,\dlp,\joerev}.  By now numerous scattering amplitudes
involving D-branes have been computed
\refs{\KleTho,\Gub,\HKleb,\Barbon, \Bachas, \Lifschytz}.

The structure of D-branes can be studied in a variety of ways.  One
type of process involves the scattering of elementary closed strings
from a D-brane \refs{\KleTho,\Gub,\HKleb,\Barbon}.  At high
energy, the strings tend to interact with virtual strings associated
with the D-brane (the `stringy halo'), and these amplitudes are very
similar to those for fundamental string-string scattering, soft with
characteristic scale the string scale.  The hard scattering 
caused by D-instantons \refs{\Green, \PolBdy} provides a tantalizing
exception.

Another possibility is to scatter one D-brane off another
\refs{\Bachas,\Lifschytz,\DanFerSun,\kp}.
In this case, if the D-branes are nonrelativistic, the effects of the
stringy halo are suppressed, and the short-distance structure of the
D-branes is visible.  In this paper we will study D$0$-brane collisions
with velocities much smaller than one but independent of $\gs$.  This
corresponds to kinetic energies and momenta up to $ \sim \ms / \gs$,
allowing very short distances to be probed.  As we shall see,
scattering in this regime produces accelerations which are small in
string units.  This means that stringy radiation is negligible, and
the effect of massive string states can be ignored, even though we are
discussing kinetic energies far above the string scale.

Besides scattering two identical branes (a D$0$-brane `colliding beam'
thought experiment), one can also scatter a low-dimensional D-brane
from a higher-dimensional D-brane (a `fixed target' experiment).  One
can think of the low-dimensional D-brane as a probe of the 
higher-dimensional D-brane's structure.

Alternatively, one can think of the low-dimensional D-brane as a probe
of the background geometry established by the higher-dimensional brane.
By now, a large body of work has developed using higher-dimensional
D-branes as elements in string compactification
\refs{\BSVDman, \gimon, \witten, \VafaF, \SenF} .
In some cases, these are dual to conventional compactifications -- for
example, the type \IIb\ D$5$-brane is S-dual to the NS $5$-brane with
a CFT description.  In others, they are to be regarded as limits in
moduli space --- for example, the type \I\ D$5$-brane is the zero size
limit of the Yang-Mills instanton.  New possibilities appear, such as
`F-theory' type \IIb\ compactifications with $7$-branes, in which the
dilaton-axion field lives in an $SL(2,\BZ)$ bundle.

In fact the two points of view are equivalent.  Polchinski \pol\
exhibited the RR charge of a D-brane by performing an open string loop
computation.  By world-sheet duality this is equivalent to classical
closed string exchange.  The relation between the D-brane description
and the geometric description of the same string compactification
ultimately rests on the same world-sheet duality.

A precise form of this relation can be established by studying a
D-brane probe \refs{\dgauge, \BanDS, \SeiI}.  Its motion can be
determined in two ways.  On the one hand, one can insert the
background field configuration in the D-brane world-volume action
directly.  On the other hand, the other D-branes in the
compactification give rise to additional open string sectors in the
probe quantum field theory, which modify its dynamics.

World-sheet duality now implies that the effects of the background
fields on the probe must be precisely reproduced by the quantum
dynamics of the probe QFT.  In the latter description, space-time
emerges as a derived concept, the low energy moduli space
of a supersymmetric gauge theory.
In some
examples the moduli space is simply related to the original space-time,
with quantum corrections to the metric.  But in general the new
space-time is a subspace of a larger configuration space.  For example,
the blowup of an orbifold singularity can be seen as the classical
minima of the potential of a linear sigma model \refs{\dm,\PolNew},
while in some multi-brane configurations, the space-time coordinates
are promoted to components of a matrix \WittenBound.

\medskip

In the present work we articulate general principles for studying this
class of problems, and use them to exhibit physical structure at
sub-stringy scales.

The two descriptions of the interaction between D-branes, the
conventional supergravity interaction mediated by massless closed
strings, and the new description using the quantum field theory of the
lightest open strings, are controlled approximations in different
domains.  Supergravity is valid at distances greater than the string
scale, while the open string theory is valid in the sub-stringy
domain.  The classical geometric picture of space-time is continuously
connected to the new short-distance picture of D-branes moving on the
moduli space of vacua (or more generally the IR configuration space)
of the world-volume field theory.

In particular, the leading long-distance behavior can be attributed to
supergravity effects, while all short-distance singularities are due
to open string effects.  Thus the proper forum for the study of short
distance structure is the quantum mechanics of open strings.

Some apparent paradoxes associated with small scales are resolved by
this observation.  For example, although the low-energy supergravity
theory predicts that the dilaton often diverges at the position of a
D-brane, this is not meaningful, because this description is invalid
there.  The correct description is in terms of a field theory of the
lightest open string modes defined in a non-singular dilaton and
metric background.  Thus string perturbation theory is applicable.

The open string moduli spaces are very similar to the closed string
geometries, and in some cases we show they are identical -- there are
no $\ap$ corrections in either description.  But, since our
description is valid at finite energy, we can also go beyond the
moduli space approximation, and study its breakdown.

Even in the open string description, the moduli spaces are often
singular when D-branes coincide.  To derive
the moduli space description we integrated out open strings
stretching between D-branes, and thus these are the usual
singularities associated with integrating out massless states in an
effective theory.

The resolution of such singularities is simply to treat the effects of
such potentially massless states more carefully.  In the context of
$0$-brane quantum mechanics, the procedure of first integrating these
states out is known as the Born-Oppenheimer approximation.  Near a
singularity of moduli space, the Born-Oppenheimer approximation will
necessarily break down, but the true quantum dynamics is non-singular.

This leads not to a modified geometrical description but rather to new
dynamical phenomena at ultra-short scales.  These include the bound
states discussed in \refs{\WittenBound, \SenBound} and elsewhere, the
resonances discussed in \refs{\DanFerSun,\kp}, and the behavior of
D$0$-brane scattering.  Our overall conclusion is that all of these
phenomena are associated with calculable sub-stringy scales.

In some problems, such as the blowup of an orbifold point, an even
more dramatic breakdown of the moduli space approximation is visible.
If the potential in the D-brane world-volume theory has a saddle point
at energy $E$, then at energies above $E$, the motion will become
qualitatively different, leading to a change of effective topology.

We begin by studying the annulus amplitude in section 2, to see the
general features of the problem.  A full treatment calls for an
effective theory combining open and closed string effects; we outline
enough of this formalism in section 8 to justify the principles stated
above.  We discuss several $0$-brane scattering problems which
illustrate our points: D$0$-D$0$ scattering in section 3, D$0$-D$4$
scattering in section 5, and D$0$ scattering off a $\BZ_2$ orbifold
fixed point in section 6.  Fine structure corrections to the spectrum
of D$0$ resonances are analyzed in section 4.  We turn to \IIb\ theory
and discuss the dyonic bound states of D$1$-branes in section 7.
Section 9 contains a summary and some discussion.

\newsec{Open string loops and closed string exchange}

The modular properties of string theory enable us to describe
interactions between D-branes in terms of open or closed string states.
If we keep all string modes both descriptions are equivalent, but a
truncation to the lightest modes of each type is valid in very different
regimes.

For D-brane separations $r>>l_s$, the interaction is most easily
described by pure massless closed string exchange.  Massive closed
strings create exponentially falling additional interactions.
Residual open string effects (to be defined more carefully later) can
be ignored.

For separations $r<<l_s$, the interaction is best described by quantum
loops of open string states.  The mass of the lightest open string
state stretching between the D-branes is $m_{W}\sim m_s (r/l_s)<< m_s$
and so dominates at small $r$.  The dynamics of this lightest state is
encoded by a ($D+1$)-dimensional quantum field theory on the brane
world-volume.  Excited open string states contribute to higher
derivative interactions in the world-volume quantum theory.  Residual
closed string effects are small at small $r$.

It is an important general property of string perturbation theory that
all potential divergences can be associated with IR effects due to
light string states.  At small $r$ the only state becoming light is
the stretched open string, so this is the only potential source of a
small $r$ singularity.

These considerations lead to the following important principle: {\it
The leading singular behavior as $r \rightarrow 0 $ of D-brane
interactions (at least to all orders in perturbation theory) is
determined by the IR behavior of brane world-volume quantum theory.}
This principle flows from a crucial connection between short distances
in space-time ($r \rightarrow 0$ ) and low energies ($m_{W}\sim m_s
(r/l_s)$) on the world-volume.

As an example of the kind of question this observation resolves,
let us turn for a moment to D$0$-brane interaction.  The classical field
configuration around a RR 0-brane has a dilaton dependence \hs\
$e^{2\phi} \sim \gs^2(1+\gs/r^7)^{3/2}$ so the effective coupling
diverges at $r=0$.  Does this mean that perturbative techniques are
invalid?  The answer is no.  In perturbation theory this problem would
manifest itself by the appearance of singular terms of the form
$(1/r^7)^{l}$ at order $\gs^l$.  But the principle tells us that these
singular terms are actually produced by the open string quantum
theory.  So if we have controlled that theory, we have controlled this
problem.

This same remark applies to metric singularities.  The classical
metric \hs\ also contains factors of $(1+\gs/r^7)^p$ and so large
curvature effects will manifest themselves by singular terms of the
form $(1/r^7)^{l}$.  Again, such singularities can only arise from the
open string quantum theory.

Singular short-distance effects normally understood as coming from
modifications of the metric or dilaton are represented by open string
dynamics.
One way to compare the two is to derive the low energy effective
action for the open string gauge theory, which will govern
the motion of the probe at low velocities.
This will include a target space metric and other couplings, which
could be compared with those derived by inserting the supergravity solution
in the probe action.

In general, the two metrics can have different short distance behavior.
The exact string theory result will cross over between
the two behaviors at the string scale.
Given enough supersymmetry, however, the leading velocity-dependent
forces turn out to be described exactly by both the supergravity
solution and the gauge theory of the lightest open strings -- neither
description receives corrections from massive string states.
We will prove this at leading order in $g_s$ for situations with
at least $1/4$
supersymmetry, such as a $0$-brane in the field of a $4$-brane.

Let us make these ideas more precise by examining the leading string
diagram contributing to the interaction between D-branes, the annulus.

\subsec{The annulus diagram -- open string channel}

At leading order in $\gs$ the interaction between two D-branes at all
distances is given by the annulus diagram with each boundary on one of
the two branes.  The velocity-dependent interaction was studied by
Bachas \Bachas\ and elaborated on by Lifschytz \Lifschytz .  We
quote their results.

Consider a $p$-brane and a $q$-brane with $p\le q$, moving at relative
velocity $v$ with impact parameter $b$.  We will consider parallel
branes.  (Non-parallel branes are a simple generalization, outlined in
appendix B.)  An open string stretched between them will have $p$ $NN$
coordinates, $8-q$ $DD$ coordinates, and $q-p$ $DN$ coordinates.  We
will refer to the numbers of coordinates of each type as $NN$, $DD$
and $DN$.  (We leave the time and the relative motion coordinates,
which have mixed boundary conditions, out of this counting.)

In the open string channel, the annulus is given by a trace over
all open string states stretched between the D-branes \joerev:
$$
{\cal A} = \int_0^{\infty}{dt \over {2t}} \sum_{i,k}
e^{-2 \pi \ap t(k^2+m_i^2)}~~.
$$
The masses $m_i$ also contain velocity-dependent effects
and at $v=0$ are given by
$m_i^2=(b^2/4\pi^2\ap^2 + N/\ap)$ where $N$ is the oscillator
excitation number minus a constant.  Doing the oscillator and
momentum sums, one finds
\eqn\oploop{
\eqalign{
{\cal A} = V_p\int_0^\infty {dt\over t} &e^{-{b^2 t/2\pi\ap}}
\left(1 \over 2\pi^2\ap t \right)^{NN/2}
\left({1\over \eta(t)}\right)^{NN}
\left({1\over \eta(t)}\right)^{DD} \cr
&\left({\theta'_{11}(0|t)\over \theta_{11}(\eps t|t)}\right)
\left({\theta_{01}(0|t)\over \eta(t)}\right)^{-DN/2} Z_F(t, \eps)
.}
}
The result for $Z_F$ depends on the case at hand.
For parallel branes with $p=q$,
\eqn\opfermp{
\eqalign{
Z_F(t, \eps) = {1\over 2\eta(t)^4}
	\sum_{ij\ne 11} e_{ij}
	 {\theta_{ij}(0|t)^3\theta_{ij}(\eps t|t)}
}
}
with $e_{00}=-e_{10}=1$, $e_{01}=-Q_1Q_2$ (the brane charges).
For parallel branes with $p\ne q$,
\eqn\opfermpneq{
\eqalign{
Z_F(t, \eps) = {1\over 2\eta(t)^4}
&\bigg[{\theta_{00}(0|t)}^{(NN+DD)/2}
	{\theta_{10}(0|t)}^{DN/2}
	\left({\theta_{00}(\eps t|t)\over\theta_{00}(0|t)}\right)\cr
  &\qquad-    {\theta_{10}(0|t)}^{(NN+DD)/2}
	{\theta_{00}(0|t)}^{DN/2}
	\left({\theta_{10}(\eps t|t)\over\theta_{10}(0|t)}\right)
\bigg]
.}
}
The parameter $\eps$ is related to the velocity $v$, $\pi \eps = {\rm
arctanh}(v)$, which for small $v$ implies $\eps \sim v/\pi$.

We can summarize \oploop\ by writing it as
\eqn\opabbrevf{\eqalign{
{\cal A} &= \int_{0}^\infty
 {dt\over t^{1+NN/2}}\ 
 {e^{-{b^2 t/ 2\pi\ap}}} f(t,\eps t) \cr
}}
or better
\eqn\opabbrevg{\eqalign{
{\cal A} &= \int_{0}^\infty
 {dt\over t^{1+NN/2}}\ 
 {e^{-{b^2 t/ 2\pi\ap}}\over\sin \pi\eps t} g(t,\eps t) .
}}
While $f$ is the modular function part of \oploop, $g$ describes the
oscillator contributions to \oploop, the modular functions with a
factor $1/\sin \pi\eps t$ (the singular part of $\theta_{11}(\eps
t|t)$) removed.  The expansion of $g$ gives a sum of
velocity-dependent terms which would appear in a quantum effective
action.  The rest of the integrand in \opabbrevg\ could be obtained by
doing the path integral for scalar particles on a trajectory ${\bf
X}(t)={\bf b} + {\bf v} X^0$.  It has singularities along the
integration contour for $\eps$ finite that give rise to an imaginary
part reflecting open string creation \Bachas.

It is convenient to introduce an effective potential ${\cal V}$ which
reproduces \opabbrevf, \opabbrevg\ when integrated along a straight line
trajectory.
\eqn\potentials{\eqalign{
{\cal A}(v,b^2) &={1 \over \sqrt{2\pi^2\ap}} \int_{-\infty}^\infty dX^0 \,\,
{\cal V}\left(v,r^2=b^2+v^2 \left(X^0\right)^2\right)\cr
{\cal V}(v,r^2) &= \int_0^\infty {dt \over t^{(1+NN)/2}} \, v e^{-r^2 t /
2 \pi \ap} f(t,\eps t) \cr
&= \int_0^\infty {dt \over t^{(1 + NN)/2}}
{v e^{-{r^2 t/ 2\pi\ap}}\over\sin \pi\eps t} g(t,\eps t)\cr}
}
For our purposes it is useful to expand in powers of $\eps \sim v$.
Let $f_n$, $g_n$ denote the function $vf$, $g$ at order
$v^{n}$ and ${\cal V}_n$ the corresponding potential:
\eqn\opabbrk{\eqalign{
{\cal V}_n &= v^n \int_{0}^\infty
 {dt\over t^{(1+NN)/2}}\ e^{-{r^2 t/ 2\pi\ap}} f_n(t)\cr
&\approx v^n \int_{0}^\infty
 {dt\over t^{(1+NN)/2}}\ {e^{-{r^2 t/ 2\pi\ap}} \over t} g_n(t) .\cr
}}
For each $n$, $g_n$ is non-singular for all
finite $t$ and behaves like $t^n$ as $t \rightarrow \infty$.

Residual supersymmetry can make leading terms in the velocity
expansion vanish.  For instance, for $p=q$ the leading contribution
to the effective action is at order $v^4$, while for $q=p+4$ the
leading contribution is at order $v^2$.

We can now appreciate the crucial point for our purposes.  Consider
the regime of short distance, $r << l_s \sim \sqap$.  It is clear from
\opabbrk\ that singular behavior as $r \rightarrow 0$ can only
come from the $t \rightarrow \infty $ part of the integration region.
This is the region well described by the lightest open string
excitations, and so all small $r$ singularities are due to stretched
open strings becoming massless.  This is the origin (at this order) of
the principle discussed earlier.

Thus we should be able to reproduce this singular behavior from the
effective field theory of the lightest open string mode.  Restricted
to this mode, the term in the effective potential corresponding to ${\cal
V}_n$ comes from a one-loop Feynman integral with $n$ external
velocity lines and $n$ internal propagators for the lightest open
string mode of mass $m^2 \sim r^2$.\foot{ The velocity couplings to
the lightest mode are perhaps most easily understood by remembering
that under T-duality velocity becomes gauge field strength
whose vector potential couples minimally. The reader
should note that here and elsewhere we occasionally suppress factors
of $l_s$ and $\ap$.}  It is
\eqn\openvel{
{\cal V}_n^{\rm lightest} \sim v^n \int {d^{p+1}k\over (k^2+m^2)^n}
	\sim v^n\left({\Lambda}^{{p+1}-2n} - m^{{p+1}-2n}\right) .
}
Here $\Lambda$ is a momentum space cutoff given by the string scale.
To compare with \opabbrk\ we rewrite this as a proper-time integral,
$$\eqalign{
{\cal V}_n^{\rm lightest} & \sim v^n \int d^{p+1}k\ \int_0^\infty
dt\ t^{n-1} e^{-t(k^2+m^2)}\cr
& \sim v^n \int_0^\infty {dt\over t^{(1+p)/2}} e^{-m^2 t} \ t^{n-1}\,.}
$$
matching the leading singular behavior of \opabbrk.

For large enough $n$, ${\cal V}_n$ is singular at small distance
(small $m \sim r$)
\eqn\gk{
{\cal V}_n \sim {v^n \over r^{2n-(p+1)}}
}
and this singularity is captured completely by the field theory
expression \openvel.  The velocity-dependent corrections have an
expansion in $v^2/r^4$.  This defines a characteristic length scale $r
\sim \sqrt{v}$, first pointed out in \Bachas, that we will discuss at
length later.

The excited open string states in \opabbrk\ make IR non-singular
contributions to ${\cal V}_n$ even at $r=0$ because they remain massive.

At finite $v$ the D-brane configuration is not supersymmetric, but
there continues to be no open string tachyon divergence in
\oploop, or equivalently by modular invariance,
no divergent sum over more and more massive closed string states.
Supersymmetric cancellations between
bosons and fermions are not exact at finite $v$ but there remains
enough asymptotic supersymmetry \KutSei\
to cancel almost all the massive closed string contributions.

When a brane interacts with an anti-brane \BanSus\ asymptotic
supersymmetry cancellations become less complete as $r$ is decreased.
At a certain $r \sim l_s$ the closed string sum diverges, or
equivalently an open string tachyon develops, signaling an interesting
annihilation instability.

In the brane-brane case, at very high velocity, $\eps \rightarrow
\infty$, the asymptotic supersymmetric cancellation of the open string
modes is badly disturbed.  This is reflected in interaction effects
whose range grows as $r \sim \sqrt{\eps}$ \Bachas.

\subsec{Closed string channel}

To go to the closed string channel, we rewrite the integral
representations \potentials, \opabbrk\ in terms of closed string
proper time $s=1/t$:
$$\eqalign{
{\cal V} &= \int_0^\infty {ds\over s^{(1+DD)/2}} v e^{-{r^2 /2\pi\ap s}}
\tilde f(s,\eps) \cr
{\cal V}_n &=v^{n}\int_0^\infty {ds\over s^{(1+DD)/2}} e^{-{r^2 /2\pi\ap s}}
 \tilde f_n(s) ~.}
$$
Now $\tilde f(s,\eps)$ is a sum over all closed string modes, weighted
by the product of the coupling constants to the two branes.  It is
computed by substituting $t=1/s$ in $f(t)$ and doing a modular
transformation to re-express this in terms of $s$.

The $s\rightarrow\infty$ limit is the closed string IR regime.
$\tilde f_n$ approaches a constant as $s \rightarrow \infty$ so the
$s^{-(1+DD)/2}$ factor gives a leading large $r$ behavior ${\cal V}_n
\sim v^{n}/r^{7-q}$, due to massless closed strings.  A term
$\exp(-m^2 s)$ in the expansion of $\tilde f_n(s)$ will integrate to
$e^{-m r}/r^{7-q}$, showing the exponential suppression of massive
closed string exchange at large $r$.
However, as $r \rightarrow 0$ the pile-up of massive closed string states will
change the behavior of the potential from ${\cal V}_n \sim
v^{n}/r^{7-q}$ to the open string result \gk, ${\cal V}_n \sim
v^{n}/r^{2n-(p+1)}$.  

In some special cases the leading term in the
velocity expansion has the same behavior in both limits.  Matching
requires $n=4-(q-p)/2$. 
Since $n$ must be even, this requires $q-p \equiv 0\ (\mod\ 4)$,
which is also the condition to have residual unbroken supersymmetry.

The simplest example is the $v^4/r^7$
potential between two $0$-branes, or more generally the $v^4$
potential between two $p$-branes.  Another example is the $v^2/r^3$
potential between a $0$-brane and a $4$-brane.  This is analogous to
the result of \dl: the one-loop prepotential in $\CN=2$, $d=4$ SYM
behaves as $F^2\log r^2$, while the massless closed string exchange
between $3$-branes in $6$ dimensions behaves the same way.

In fact, in these special cases the sum over string states degenerates to the
lightest states: either open strings with no oscillator excitations,
or massless closed strings.  We show this explicitly in appendix A.

\subsec{Decomposing the annulus moduli space}

We now address the important question of fitting together the regimes
discussed above.  We can rewrite \opabbrk\ in a form which allows
controlling the two limits of small and large $r$.  We do this by
dividing the modular integral into an open string IR region (large
$t$) and a closed string IR region (small $t$ or large $s$).  We
choose a cutoff parameter $T_0$ and assign $t>T_0$ to the open string
region and $t<T_0$ to the closed string region.  We can then write
\eqn\opdecomp{\eqalign{
{\cal V}_n &= v^{n}\int_{T_0}^\infty
 {dt\over t^{(1+NN)/2}}\ e^{-{r^2 t/ 2\pi\ap}} f_n(t) \cr
&+ v^{n}\int_{1/T_0}^\infty
 {ds\over s^{(1+DD)/2}}\ e^{-{r^2 / 2\pi\ap s}} \tilde f_n(s) .
}}

The large $r$ behavior is, as discussed above, dominated by the closed
string region $t<T_0$.  The open string region's effect at large $r$
is easily estimated.  For $t>T_0$, the integrand is exponentially
suppressed by at least $\exp( -r^2 T_0/2\pi\ap)$. ( We can estimate
$f_n(t)$ by $f_n(T_0)$ which is bounded.) So the residual open string
effects are exponentially small.

The small $r$ behavior is dominated by the open string region.  We can
now estimate the closed string region's effect at small $r$.  Let us
assume for simplicity that $DD \geq 2$, so that 
${\cal V}_n$ decreases at large $r$.\foot{
This is true unless $p$-branes with $p\ge 7$ are involved, in which
case the massless closed string propagator is IR divergent.  The
simplest way to treat this case is to add another source of opposite
charge.  The combined ${\cal V}_n$ will decrease at large $r$, and the same
discussion holds.  } Then the $s$ integral in \opdecomp\ converges at
$r=0$, and has a Taylor expansion in $r$.  So the closed string region
contributes a smooth non-singular interaction at small $r$.  The
characteristic scale of variation for this interaction is of course
the string scale.

This confirms in detail at this order the principle discussed earlier.
Closed string effects are soft at small $r$.  Heuristically one can
say that closed string excitations like the metric and dilaton are
smeared out at string scale.\foot{This point has also been emphasized
by Joe Polchinski.}  Singular short-distance effects normally
understood as coming from modifications of the metric or dilaton are
represented by open string dynamics.

To extend this discussion to all orders in string perturbation theory
we need a decomposition of the full moduli space which makes
factorization on all open and closed string intermediate states
manifest.  Such a decomposition has been developed by Zwiebach
\zwiebach, and having this will make the extension straightforward.
Since the details of the argument are not required for the examples
studied in this work, we defer it to section 8.

\newsec{0-brane scattering}

At this point we have justified analyzing 0-brane dynamics at short
distances in terms of the lightest modes of the open strings which end
on the branes.  In this section we use this description to study
scattering of two 0-branes, extending the work of \refs{\DanFerSun, \kp} .

We first formulate the system we wish to analyze.  The relevant
degrees of freedom are the dimensional reduction of an $\CN=1$
supersymmetric $U(2)$ vector multiplet from $9+1$ dimensions to $0+1$
dimensions.  On the world-line of the brane we have the fields
$$
\eqalign{
A_0 &= {i \over 2} \left(A_0^0 \identity + A_0^a \sigma^a\right)\cr
\phi_i &= {i \over 2} \left(\phi_i^0 \identity +
       \phi_i^a \sigma^a\right)\cr
\psi &= {i \over 2} \left(\psi^0 \identity +
       \psi^a \sigma^a\right)\cr}
$$
where $A_0$ is a 0+1 dimensional $U(2)$ gauge field, $\phi_i$
($i=1,\ldots,9$) is a collection of adjoint Higgs fields, $\psi$ is a
sixteen component adjoint spinor (Majorana-Weyl in $9+1$), and $a$ is
an adjoint $SU(2)$ index.

The $U(1)$ Higgs fields $\phi_i^0$ are center of mass coordinates 
(in string units) for
the two 0-branes.  A non-zero expectation value for the $SU(2)$ Higgs
fields $\phi_i^a$ breaks $SU(2)$ down to $U(1)$; the massless degrees
of freedom associated with the unbroken $U(1)$ act as relative position
coordinates for the two 0-branes.

At low velocity, we expect these degrees of freedom to be governed by
the dimensional reduction of the super-Yang-Mills action, namely
\eqn\YM{
S = \int dt \, {1 \over 2 \gs} {\rm Tr} \, F_{\mu \nu} F^{\mu \nu}
            - i \, {\rm Tr} \, \bar\psi \, \Gamma^\mu D_\mu \psi\,.
}
where
$$\eqalign{
F_{0i} &= \partial_0 \phi_i + [A_0,\phi_i]\cr
F_{ij} &= [\phi_i,\phi_j]\cr
\noalign{\smallskip}
D_0 \psi &= \partial_0 \psi + [A_0,\psi]\cr
D_i \psi &= [\phi_i,\psi]\,.\cr}
$$

The characteristic length and  energy 
scales of the low-lying resonances of this system 
can be determined by a scaling analysis
\refs{\DanFerSun, \kp}.  
The entire $\gs$ dependence of the action can be eliminated by
introducing the following scaled quantities:
\eqn\scaledvars{
t = \gs^{-1/3} t_{11} \qquad A_0 = \gs^{1/3} A_0^{11} \qquad
\phi_i = \gs^{1/3} \phi_i^{11}}
so that $F_{\mu\nu} = \gs^{2/3} F_{\mu\nu}^{11}$ and $S=\int dt_{11}
(F^{11})^2$ .

So we see \refs{\DanFerSun, \kp}  that the
characteristic size of the low-lying resonances
is the eleven-dimensional Planck length $\lp11 = \gs^{1/3} \ls$,
much less than the string length at weak coupling,  and
the characteristic energy is $\gs^{1/3} \ms$, which corresponds to a
velocity $v \sim \gs^{2/3}$, small at small $\gs$.  (Remember
that the 0-brane mass is $m_0 \sim m_s/g_s$. Note also that, despite
its name, $t_{11}$ is {\it not} the time measured in 
eleven-dimensional Planck units).

This scaling analysis can be summarized by noting
that the coupling $\gs$ has world-line dimensions of $m^3$ and $m \sim
r$ so the natural length scale is $\gs^{1/3}\ls$.

Corrections to this action come in two types.  First, there are
corrections which involve higher powers of the field strength but no
covariant derivatives.  These can be re-summed into the
Dirac--Born--Infeld form \Lei
$$
S_{\rm DBI} = - {1 \over \gs} \int {\rm Tr} \,\,
\sqrt{- \det \left(\eta_{\mu\nu} \identity + F_{\mu\nu}\right)}\,.
$$
At low velocities, corresponding to weak field strengths, this reduces
to the Yang-Mills action above.
The size of these corrections will be discussed
in the next section.   A second type of correction involves
covariant derivatives of the field strength, that is, the acceleration
and higher time derivatives of the brane velocity.

For our analysis just in terms of the Yang-Mills action to be valid,
we must keep both the velocity and the acceleration of the branes
small during the scattering event.  We analyze this qualitatively
below.

\subsec{Qualitative analysis}

To gain some insight into the finite velocity
dynamics we study a simplified model
which captures most of the essential physics of the full problem.  We
introduce a two-component wavefunction $\Psi(x_1,x_2,w_1,w_2)$, and take
the Hamiltonian
$$
H = - \half \gs \left(\nabla^2_x + \nabla^2_w \right) + {1 \over 2 \gs}
x^2 w^2 + \left[\matrix{& 0 & x_1 - i x_2 \cr & x_1 + i x_2 & 0 \cr}
\right] \,.
$$
An identical toy model was introduced in \DeWitLuNi.  The fields $x_1$,
$x_2$ are massless bosons.  They correspond to the Higgs fields which
break $SU(2)$ down to $U(1)$, and should be thought of as relative
position coordinates for the two 0-branes.  The fields $w_1$, $w_2$
have a mass $\vert x \vert$, which can be thought of as arising from
$SU(2)$ breaking.  They represent unexcited open strings stretched
between the two 0-branes.  Finally, the two components of the
wavefunction represent the Fermi superpartners of $w_1$, $w_2$.  The
degrees of freedom present in the full 0-brane quantum mechanics
problem correspond to eight copies of the simplified model.

We assume that the incoming state, with velocity $v$ and impact
parameter $b$, is sharply localized about the expectation values
$<x_1(t)> = v t$, $<x_2(t)> = b$.  We wish to find how the massive degrees
of freedom influence the free propagation of this wavepacket.

In a Born-Oppenheimer approximation, we treat the massless fields
$x_i$ as slowly varying, and quantize the massive degrees of freedom
as if the background were static.  The massive bosons then have energy
levels $\sqrt{v^2 t^2 + b^2} \left(n_1+n_2+1\right)$, while the
fermions have energy levels $\pm \sqrt{v^2 t^2 + b^2}$.  Initially, we
put the massive degrees of freedom in their ground state where, in a
caricature of unbroken supersymmetry, the zero-point energies cancel.

The Born-Oppenheimer approximation will be valid as long as the
frequency of the massive oscillators is slowly varying.  The criterion
is that the change in period during a single oscillation should be
small, or equivalently
$$
{\delta \omega \over \omega} = {\dot{\omega} \omega^{-1} \over \omega}
\approx {v \over x^2} < 1 \,.
$$
This defines a region, $\vert x \vert > v^{1/2}$, in which the
Born-Oppenheimer approximation is valid, and in which the massive
degrees of freedom stay in the ground state of their slowly varying
Hamiltonian.  The importance of the separation scale $v^{1/2} \ls$ was
first noted by Bachas \Bachas.

Eventually, provided the impact parameter $b$ is less than $v^{1/2}$,
the 0-branes enter the region $\vert x \vert < v^{1/2}$, which we
refer to as the `stadium'.  In this region the Born-Oppenheimer
approximation breaks down, and we must take all the `non-Abelian'
degrees of freedom into account.  Consider the evolution of the Bose
degrees of freedom.  In the Born-Oppenheimer approximation, the Bose
wavefunction is a Gaussian, with spread in the non-Abelian directions
$\Delta w \sim (\gs/x)^{1/2}$.  So, when entering the stadium, the
wavefunction is a Gaussian with width $\gs^{1/2} / v^{1/4}$.  If we
neglect the potential, the wavefunction will spread by free diffusion
in crossing the stadium.  During the time $\sim v^{-1/2}$ that the
0-branes remain in the stadium, the amount of additional spreading
turns out to be of the same order as the initial spread $\gs^{1/2} /
v^{1/4}$ -- narrow enough that it is self-consistent to neglect the
potential.  The evolution of the Fermi degrees of freedom is also
simple: by comparing the gap between the two energy levels (at most
$v^{1/2}$) with the time $\sim v^{-1/2}$ spent in the stadium, we see
that the Fermi wavefunction does not have enough time to evolve
significantly.

This shows that, while in the stadium, the wavefunction does not
change much.  As a first approximation, we take it to be
constant.

Finally, the 0-branes leave the stadium.  At this point we project the
massive degrees of freedom onto their new ground state, to find the
outgoing wavefunction for the $x_i$.  The overlap of the massive
bosons with their new ground state is unity, because the bosonic
ground state is the same on both sides of the stadium (note that the
bosonic Hamiltonian is invariant under $t \rightarrow -t$).  For the
fermions, on the other hand, the Hamiltonian changes significantly in
crossing the stadium: at small $b$, the ground state on one side of
the stadium resembles the excited state on the other side, and
vice-versa.  In effect the ground and excited states are interchanged
in crossing the stadium.  This means that the overlap of the fermion
wavefunction with the new ground state vanishes as the impact
parameter goes to zero.  Some straightforward algebra shows that, at
this level of approximation, the outgoing state is proportional to
$\vert b \vert / v^{1/2}$.

A striking feature of this system is that scattering at small enough
impact parameter ($b < v^{1/2}$) is almost always accompanied by the
excitation of a massive fermion.  Translated into D-brane language,
this means that scattering at small impact parameter will almost
always produce straight fermionic open strings stretched between the
0-branes.

This gives us a qualitative picture of the dynamics.  In a scattering
experiment at velocity $v$, the most likely impact parameter is of
order the size of the stadium, $b \sim v^{1/2}$.  Scattering at this
impact parameter will typically produce some number of unexcited open
strings,
ranging from zero up to the maximum of sixteen open strings set by the
exclusion principle.
If no open strings are created 
the 0-branes leave the stadium in an elastic event.
If open strings are created,  the 0-branes are
subject to a linear potential from the stretched strings, with slope
$\sim 1/\ap$.  At moderate velocities, the 0-branes have an enormous
kinetic energy compared to the string scale, so they can travel a long
distance before they are brought to rest.  {}From energy and angular
momentum conservation one sees that they reach a maximum separation
$\sim v^2 \ls / \gs$, and are deflected by a small angle $\theta
\sim \gs / v^{3/2}$ from their initial trajectory.  This scattering angle
corresponds to a momentum transfer $\sim \ms / \sqrt{v}$.

Eventually, the 0-branes are pulled back in and collide again.  The
fermion ground and excited states are interchanged for a second time,
and after the second collision, the 0-branes separate, typically with
some different combination of open strings present.  Repeated
collisions can take place, until eventually either a collision happens
to produce no open strings, or all the open strings decay between one
collision and the next.  These are highly excited examples
of the phenomenon,
first discussed in \DanFerSun,
of resonances in the linear
potential between 0-branes.

To show that multiple collisions are possible, we need to show that
the probability for all the open strings to annihilate between one
collision and the next is small.  We estimate this probability as
follows.  The amplitude for two pieces of string to annihilate is
proportional to $\gs$.  Different parts of the stretched strings
contribute incoherently to the total annihilation rate, so the
probability of annihilation is
$$\eqalign{
&\gs^2 \times \hbox{\rm (length of stretched string)} \times \hbox{\rm (time
between collisions)} \cr
\sim \, &\gs^2 \times {v^2 \over \gs} \times {v \over \gs} \cr
= \, &v^3\,.\cr}
$$
The probability, although not suppressed by the string coupling, is
small provided the velocities are small.

This means that the resonances have a long lifetime, of order the time
between collisions $v \ls /\gs$, corresponding to a width $\Gamma \sim
\gs \ms /v$.  The resonance energies are fixed by the semiclassical
quantization condition for a linear potential, $E_n \sim \gs^{1/3}
n^{2/3} \ms$ \DanFerSun.  Note that, as expected for a semiclassical
trajectory making a single orbit, the width of a resonance is of order
the spacing between successive resonances.  These resonances would be
seen as a rapid variation in the smooth scattering amplitude we find
in the next section.

Other effects that potentially could alter this picture include 
massless closed string radiation and excitation of the very light open string
modes that exist because the stretched string is so long.
In fact both of these effects are small.  As an example, consider
RR photon radiation from the moving 0-branes. Radiation power
emitted from a non-relativistic charge goes as $a^2$ where $a$ is the 
acceleration.  The 0-brane has mass $\sim m_s/g_s$ and experiences a
force from the stretched string of order one  so $a \sim \gs$.  The
total energy emitted in a cycle is then $\sim a^2 v/\gs \sim v \gs$
which is very small, and in particular is much less than the spacing
between resonances.  In fact, the wavelength of this radiation is large
compared to the separation between the 0-branes, so interference effects
(the absence of dipole radiation) suppress this effect further.

The energy transferred to the light open 
string modes can be estimated by modelling the 0-branes as `moving
mirrors' coupled to the stretched string quantum field.
The Casimir force between the two mirrors is velocity-suppressed
due to supersymmetry and falls off at large $r$.
Radiation
from a non-relativistic moving mirror \birdav\
goes as $a^2$ 
and so the above estimate applies.

Two important limitations on the velocity must be satisfied for our
analysis to be valid.  A lower limit comes from the uncertainty
principle.  During the time $\sim v^{-1/2}$ the 0-branes spend in the
stadium, their wavefunctions will spread a distance $\sim \gs^{1/2} /
v^{1/4}$.  We must have $v > \gs^{2/3}$, the characteristic velocity
of the low-lying resonances, or the wavepackets will
diffuse an uncontrollable amount.  This means, in particular, that our
analysis is not sensitive enough to study the bound state at
threshold or the low-lying resonances in detail.

An upper limit on velocity comes from demanding that the corrections
to the Yang-Mills action be small.  Recall that these corrections come
in two types.  One type involves higher powers of the velocity, and
can be neglected provided the velocity is small.  Another type
involves time derivatives of velocity (measured in string units).
These corrections are negligible provided that the velocity is small
and the collision takes a long time in string units.  We identify the
duration of the collision $\sim v^{-1/2} \ls$ with the time to cross
the stadium.  So this is long and
corrections are negligible provided $v$ is
small.

So we are restricted to the non-relativistic regime.\foot{This is a
wider range of velocities than was proposed in \kp.}  But because
0-branes have a mass $m_0 \sim \ms / \gs$, small but finite velocities
correspond to momenta and energies $\sim \ms / \gs$ far above the
string scale.  The resonances discussed above persist throughout this
regime.  They provide direct evidence for variations in the 0-brane
wavefunction over lengths as short as $\R11$.

Our approximations break down when the 0-branes become relativistic,
and the length scales probed are shorter than $\R11$.  Exploring
further into this region may well teach us much about M-theory dynamics.

\subsec{Hard scattering}

High energy scattering at fixed angles is a classic physical probe of
the short-distance structure of a theory.  In this section we consider
the amplitude for fixed-angle scattering of two 0-branes, adapting the
string results obtained by Bachas \Bachas~to the quantum mechanics
problem.

Based on the qualitative treatment above, we expect our analysis to be
valid only for a certain range of velocities.  A minimum velocity,
$v > \gs^{2/3}$, is necessary so that the 0-branes can be well-localized
during the scattering event and simple approximation techniques
brought to bear.  A maximum velocity, $v < 1$, is
necessary to keep the 0-branes non-relativistic, so that corrections
to the Yang-Mills action are small.

We expand the Yang-Mills action around a background corresponding to
motion in a straight line.
$$\eqalign{
<\phi_1> &= (0,0,v t) \cr
<\phi_2> &= (0,0,b) \cr}
$$
This expansion is the eikonal approximation, valid when $k \lp11 >>1$.

The background breaks $SU(2)$ down to $U(1)$.  We work in background
field gauge, and integrate out the massive degrees of freedom at one
loop.  This gives rise to the determinants (in Euclidean space, with
$\tau = i t$ and $\gamma = -iv$)
$$\eqalign{
& \det{}^{-6} \left(-\partial_\tau^2 + \gamma^2 \tau^2 + b^2\right)
\det{}^{-1} \left(-\partial_\tau^2 + \gamma^2 \tau^2 + b^2 + 2
     \gamma\right) \cr
& \det{}^{-1} \left(-\partial_\tau^2 + \gamma^2 \tau^2 + b^2 - 2
     \gamma\right)
\det{}^8 \left[\matrix{& \partial_\tau & \gamma \tau - i b \cr
       & \gamma \tau + i b & \partial_\tau \cr}\right]\,. \cr }
$$
The phase shift as a function of impact parameter and velocity can be
expressed using a proper-time representation for the determinants.
\eqn\phase{
\delta(b,v) = - {1 \over 4} \int_0^\infty {ds \over s} \, e^{-s b^2}
{1 \over \sin sv} (16 \cos sv - 4 \cos 2sv - 12)
}
This result can be extracted from the work of Bachas \Bachas\
by truncating his string calculation to
the lightest open string modes.  The outgoing
wavefunction is a plane wave modulated by a factor
$$
e^{i \delta(b,v)} = \tanh^4 \left(\pi b^2 \over 2 v\right) {i b^2 + v \over
i b^2 - v} \left[\Gamma\left(-{i b^2 \over 2 v}\right) \Gamma\left(\half +
{i b^2 \over 2 v}\right) \over \Gamma\left({i b^2 \over 2 v}\right)
\Gamma\left(\half - {i b^2 \over 2 v}\right) \right]^4 \,.
$$

As $b \rightarrow 0$ the wavefunction vanishes like $b^8$, precisely
the behavior we found in our qualitative analysis of the previous
section (remember that the full quantum mechanics problem corresponds
to eight copies of the simplified model).  This agreement shows that a
one-loop approximation is reasonably accurate in the velocity range
$\gs^{2/3} \ll v \ll 1$, even at small $b$.  The physical interpretation
is that the wavefunction goes to zero because scattering at small
impact parameter is almost always accompanied by inelastic production
of fermionic open strings stretched between the 0-branes.

The amplitude for elastic scattering with momentum transfer $q$ is a
Fourier transform of the outgoing wavefunction.  At large momentum
transfer, the Fourier integral is dominated by the poles closest to
the real axis, and the elastic scattering amplitude is exponentially
suppressed.
$$
f(q,v) = \int d^8 b e^{i q \cdot b} \left(e^{i \delta(b,v)} - 1\right)
\sim e^{- q \sqrt{v/2}}
$$

Upon expressing the scattering amplitude in terms of the initial
momentum $k = v/\gs$ and scattering angle $\sin \theta/2 = q/2k $
, we see that the eleven-dimensional Planck length sets the scale
for the momentum dependence.
\eqn\hardscat{
f(k,\theta) \sim e^{-\sqrt{2} \sin(\theta/2) \left(k \lp11
\right)^{3/2}}
}
0-brane scattering at fixed angles can directly measure the 
eleven-dimensional Planck length.  The fact that the amplitude is
exponentially suppressed at momentum transfers that are large compared
to the eleven-dimensional Planck scale gives some hope that 0-branes
will not ruin the good high-energy behavior displayed by perturbative
string theory.

Relativistic corrections will certainly become important 
as $v \rightarrow 1,~ k \rightarrow 1/R_{11}$, 
signalling this additional scale.  Bachas \Bachas\ has included
these effects at leading order in $\gs$ and finds a growing stringy 
halo developing of size $\sim \sqrt{\log(k R_{11})}$.  It is an
important question
whether perturbation theory remains accurate in this region.

The scattering amplitude can also be written in terms of the velocity
and scattering angle.
$$
f(\theta,v) \sim e^{- \sqrt{2} \sin(\theta/2) v^{3/2} / \gs}
$$

As $v \rightarrow 1$ hard scattering becomes an ${\cal O}(e^{-1/\gs})$
effect, comparable to the strength of non-perturbative effects in
string theory \ShenkerNP.  In particular, a relativistic 0-brane has
enough momentum to be able to resolve the radius of the eleventh
dimension.  This is another way of seeing that our analysis, in terms
of perturbative string states attached to the 0-branes, may break down
as the 0-branes become relativistic.

Within the quantum mechanical context, 
we can estimate higher order corrections to the eikonal approximation.
These corrections are suppressed by inverse powers of ~$(k \lp11) >> 1$
and have the same exponential falloff as the leading contribution
\hardscat . So the calculation we have presented is reliable in the
range we have discussed.

\newsec{Fine structure and  $\R11$}

In the previous section, we were only able to study non-relativistic
0-brane scattering, in which the 0-branes do not have enough momentum
to be able to directly resolve the radius of the eleventh dimension
$\R11 = \gs \ls$.  In this section, we will present indirect 
indications for the role of $\R11$, by studying the effect of
corrections to the Yang-Mills action \YM.

As discussed in the previous section, 
0-brane scattering displays a series of resonances
\refs{\DanFerSun, \kp}.  The characteristic size of the low-lying
resonances
is the eleven-dimensional Planck length $\lp11 = \gs^{1/3} \ls$, and
the characteristic energy is $\gs^{1/3} \ms$, which corresponds to a
velocity $v \sim \gs^{2/3}$.  At leading order, the resonance is
controlled by the Yang-Mills action.

We want to determine how corrections to the Yang-Mills action affect
the resonance energies.  In particular, we consider corrections which
involve higher powers of the field strength.  These can be re-summed
into the Dirac--Born--Infeld action \Lei
$$
S_{\rm DBI} = - {1 \over \gs} \int dt \, {\rm Tr} \, \sqrt{-\det
\left(\eta_{\mu\nu} \identity + F_{\mu\nu}\right)}\,.
$$
Recall that we are measuring distance and time in string units.  To
study the effect of higher order terms in the DBI action, it is
convenient to go to eleven dimensional scaled variables
as in \scaledvars\  .

The action takes the form
$$
S_{\rm DBI} = - {1 \over \gs^{4/3}} \int dt_{11} \, {\rm Tr}
\sqrt{- \det \left( \eta_{\mu\nu} \identity + \gs^{2/3} F_{\mu\nu}^{11}
\right)}\,.
$$
Expanding in powers of $\gs$, the first non-trivial term is ${\cal
O}(1)$, which shows that the kinetic energy of the resonance is indeed of
order one (in units of $\gs^{1/3} \ms$).  There are higher order
corrections to this result, which will shift the energy by an amount
of order $\gs^{4/3}$ (again in units of $\gs^{1/3} \ms$).

This is all directly analogous to atomic physics.  Recall that the
electron in a hydrogen atom moves with a characteristic velocity set
by the fine structure constant, $v = \alpha$.  The characteristic size
of the atom is the Bohr radius, $a = {1 \over \alpha m_e}$, and the
characteristic scale for the energy levels is $\alpha^2 m_e$.
Relativistic corrections to the spectrum produce fine structure
splittings of order $\alpha^4 m_e$ in the unperturbed energy levels.
These corrections (in particular the Darwin term \Sakurai) are
indirect evidence for the existence of a new short length scale: the
electron Compton wavelength $\lambda = 1/m_e$, smaller than the Bohr
radius by a factor of $\alpha$.  Upon identifying
the electron mass and velocity with the 0-brane mass and velocity,
we see that the 0-brane resonances behave just like a hydrogen atom:
\bigskip
\centerline{\vbox{\offinterlineskip
\hrule
\halign{&\vrule#&
	\strut\quad#\quad\cr
height3pt&\omit&&\omit&\cr
&Electron\hfil &&0-brane\hfil & \cr
height3pt&\omit&&\omit&\cr
\noalign{\hrule} 
height3pt&\omit&&\omit&\cr
& mass $m_e$ \hfil && mass $m_0 = \ms/\gs$\hfil &\cr
& velocity $\alpha$\hfil && velocity $v_0 = \gs^{2/3}$\hfil & \cr
&Bohr radius $\sim 1/\alpha m_e$ && 
size of resonance $1/m_0 v_0 \sim \gs^{1/3} \ls \sim \lp11$ & \cr
& energy levels $\sim \alpha^2 m_e$ && 
resonance energy $\sim m_0 v_0^2 \sim \gs^{1/3} \ms$ & \cr
&fine structure $\sim \alpha^4 m_e$ && 
energy shifts $\sim m_0 v_0^4
\sim \gs^{5/3} \ms$& \cr
&Compton wavelength $1/m_e$ && 
Compton wavelength $1/m_0 \sim \gs \ls \sim \R11$ & \cr
height3pt&\omit&&\omit&\cr}
\hrule} }
\medskip\noindent
This analogy suggests that the fine structure shifts in the spectrum
of 0-brane resonances should be interpreted as indirect evidence for the
existence of a new length scale: the Compton wavelength of a 0-brane,
equal to the radius of the eleventh dimension.  Of course the 0-brane
resonances are much broader than atomic levels and so this fine
structure will be harder to discern.

The DBI action has the same form as the action for a relativistic
point particle, so it is not surprising that the atomic physics
analogy works so well.  There are other corrections, such as terms
involving covariant derivatives of the field strength, which are not
included in the DBI action, but the leading corrections seem to come
from DBI.  For example, a term in the action ${1 \over \gs} \int
(\partial F)^2$, which would produce an order $\gs \ms$ shift in
energy, is absent from the open superstring effective action \AndTsey.

\newsec{D$0$--D$4$ brane scattering and metrics on moduli space}

In this section we consider scattering a $0$-brane from a $4$-brane.
The main novelty, compared to D$0$--D$0$ scattering, is that the
reduced amount of supersymmetry allows a non-trivial metric on moduli
space.

We first construct the appropriate 0-brane quantum mechanics.  Besides
the $0$-$0$ strings, which give rise to massless world-line degrees of
freedom, there are light modes from the $0$-$4$ strings, described in
\dgauge.  It is convenient to describe these degrees of freedom in
terms of the reduction of an $\CN=1$ $d=6$ Abelian gauge theory to
$0+1$ dimensions.  In this language, the $0$-$0$ strings give rise to
a vector multiplet and a neutral (adjoint) hypermultiplet.  The
$0$-$4$ strings give rise to an additional hypermultiplet, charged
under the $0$-brane $U(1)$.

We place the $4$-brane in the $06789$ plane and scatter a $0$-brane
from it.  As in our analysis of section 3.2, we turn on a background
on the $0$-brane corresponding to motion in a straight line with
velocity $v$ and impact parameter $b$.  Any motion of the $0$-brane
parallel to the $4$-brane is trivial and we ignore it.  This
background gives a mass to the $0$-$4$ hypermultiplet, which we
integrate out at one loop.  This gives rise to the determinants
($\gamma = -i v$)
$$
\det{}^{-2} \left(-\partial_\tau^2 + \gamma^2 \tau^2 + b^2\right)
\det{}^2 \left[\matrix{& \partial_\tau & \gamma \tau - i b \cr
       & \gamma \tau + i b & \partial_\tau \cr}\right]\,.
$$
The eikonal phase shift can be expressed using a proper-time
representation for the determinants:
\eqn\eik{\eqalign{
\delta(b,v) &= \int_0^\infty {ds \over s} e^{-s b^2} \tan {sv \over 2} \cr
e^{i \delta(b,v)} &= - i \, \tanh\left({\pi b^2 \over 2 v }\right)
{\Gamma\left(-{i b^2 \over 2 v}\right)
\Gamma\left({1 \over 2} + {i b^2 \over 2 v}\right) \over 
\Gamma\left({i b^2 \over 2 v}\right)
\Gamma\left({1 \over 2} - {i b^2 \over 2 v}\right)} . \cr} }
This can be extracted from the results given in \Lifschytz.

This expression makes it clear that D$0$--D$4$ scattering is
qualitatively similar to D$0$--D$0$ scattering.  That is, the
$0$-brane and $4$-brane interact in a stadium of size $\sim v^{1/2}$,
significant inelasticity is present at small impact parameter, and
hard scattering is exponentially suppressed, falling off as
$e^{-\sqrt{2} \sin(\theta/2) (k \lp11)^{3/2}}$.

The novel feature of the D$0$--D$4$ system is that the equivalent of
$\CN=1$ $d=6$ supersymmetry allows a non-trivial metric on the D$0$
moduli space.  The effective action determined by \eik\ is
\eqn\effacc{
\eqalign{
S_{\rm eff} &= \int dt \, \left\lbrace {1 \over 2 \gs} v^2 +
v \int _0^\infty {ds \over \sqrt{\pi s}} \, e^{-s r^2} \tan 
{sv \over 2} \right\rbrace \cr
&= {1 \over 2 \gs} \int dt \,  \left(1 + {\gs \over 2 r^3}\right) v^2 +
{\cal O} \left(v^4/r^7\right) \cr}}
so there is a metric on moduli space
\eqn\metric{
ds^2 = \left(1+{\gs \over 2 r^3}\right) \left(dr^2 + r^2 d\Omega_4^2\right)
}
where $d\Omega_4^2$ is the metric on the unit 4-sphere.

The D$0$-D$4$ system is special, in the sense of section 2.2, in that
the quantum mechanics of the lightest open strings exactly reproduces
the $1/r^3$ fall off of the metric at large $r$, which one normally
thinks of as due to massless closed string exchange.  In fact, the
metric \metric\ is likely to be exact.  It receives no $\ap$
corrections, as shown in appendix~A.2.  The non-renormalization
theorems for $\CN=2$ $d=4$ supersymmetry, assuming they are valid
after dimensional reduction, imply that there are no perturbative
$\gs$ corrections \Superspace.  Finally, non-perturbative corrections
must come from instantons preserving half the supersymmetry, and there
is no candidate instanton in this problem.

To visualize the moduli space geometry, it is convenient to change to
a new radial coordinate $\rho^2 = r^2 (1 + \gs / 2 r^3)$, so that the
metric has the form $ds^2 = f(\rho) d\rho^2 + \rho^2 d\Omega_4^2$.
The moduli space can then be embedded as a hypersurface in the $z>0$
half of a flat six-dimensional space, with metric $ds^2 = d\rho^2 +
\rho^2 d \Omega_4^2 + dz^2$.  The embedding is given by setting $\rho
= \rho(z)$, where
$$
\rho(z) \sim \cases{
4 \gs / z^2 & as $z \rightarrow 0$ \cr
z / \sqrt{3} & as $z \rightarrow \infty$ . }
$$
The asymptotically flat space far from the 4-brane (at small $z$) is connected
by a throat, with size of order $\lp11$, to an asymptotically conical
space at small $r$ (large $z$).  The point $r=0$ is at infinite
distance.

There are several important features of this metric.  The corrections
to flat space are suppressed by $\gs$, and are singular at small $r$.
Neither of these properties holds for the moduli space metrics of
traditional ``fat'' solitons (e.g., the Atiyah-Hitchin metric).  These
are further indicators of the small size of these objects.  Moreover,
the corrections to flatness become important when $r \sim \gs^{1/3}
\ls \sim \lp11$, another indication of the basic role of this scale in
D0-brane dynamics.  

By considering the M-theory origin of these objects, one can see that
the D$0$ -- D$4$ system must have a bound state at threshold.  The
D$4$-brane is the simultaneous spacetime and world-volume dimensional
reduction of an M-theory 5-brane.  Kaluza-Klein states of the 5-brane
world-volume theory are BPS saturated, with a charge under the gauge
field arising from $g_{\mu \, 11}$.  The only candidate for this
object in the string theory is a 0-brane -- 4-brane bound state at
threshold.

We can count the degeneracy of these states from the M-theory point of
view.  The 5-brane itself breaks half of the eleven-dimensional
supersymmetry, so its massless world-volume fields are eight bosons and
eight complex fermions \refs{\KapMich, \BergRooOrtin}.
Its ground state is therefore $2^8$-fold degenerate, and
Kaluza-Klein excitations provide $16$ states for each unit of
Kaluza-Klein momentum, which corresponds to 0-brane charge.
This
matches the expectation from the D-brane point of view: as the
combined D$0$-D$4$ system preserves only 1/4 of the ten-dimensional
\IIa\ supersymmetry, its ground states are $2^{12}$-fold degenerate.

It is convenient to
decompose the $0$-$0$ open string fermion zero
modes into those from the $\CN=1$ $d=6$ vector
multiplet (a spinor transverse to the $4$-brane)
and a neutral hypermultiplet (a longitudinal spinor).
Quantizing the hypermultiplet and the $4$-$4$  
zero modes already provides $2^{12}$ states,
so the correct result will be reproduced if the remaining
supersymmetric quantum mechanics has a single zero energy bound state.  

This system was recently studied by Sethi and Stern \SethiStern, 
who argued that its bound states describe H-monopoles in toroidal
compactification of heterotic string theory \GauntHarv.
They computed the index to be $1$, and thus established the existence
of a bound state.

We will study the bound state in the moduli space approximation,
as a normalizable
zero mode of the Laplacian on moduli space (much as in
\refs{\GibMan,\Senbdmon}).
Since we integrate out the $0$-$4$ open strings to derive this description,
they no longer contribute zero modes.
Quantizing the $0$-$0$ vector multiplet leads to a sum of
odd $p$-forms on moduli space, and the bound state could be any of these.

Zero modes do exist, and can be constructed as follows.  Consider the
0-form $\omega_0 = \left(1+\gs/2r^3\right)^{-1/2}$.  It is annihilated
by the Laplacian, but is not normalizable.  Two more harmonic forms, a
one-form and its dual 4-form, can be constructed from $\omega_0$ by
taking derivatives.
\eqn\bswave{\eqalign{
\omega_1 &= d \omega_0 = {dr \over r^4 \left(1 + \gs/2r^3\right)^{3/2}} \cr
\omega_4 &= {}^* d \omega_0 \cr}
}
Both $\omega_1$ and $\omega_4$ have finite norm; they are the only
normalizable harmonic forms on moduli space.  

Since the moduli space approximation becomes arbitrarily good at large
$r$, and the zero energy bound state wave function falls off as a power, 
it must have this large $r$ behavior.
It is concentrated on the throat of the moduli space, at
$r \sim \gs^{1/3}$, so as expected the bound state has a
size $\sim \lp11$.

The moduli space approximation breaks down at small $r$.  The natural
length scale for this breakdown is the stadium size
$b \sim \sqrt{v} \sim \gs^{1/3}$ where the
Born-Oppenheimer approximation discussed in section 3.1
ceases to be valid.
This is signaled by the
higher order terms in \effacc\
becoming important.  Below this scale, the system should be studied
by retaining all the degrees of freedom.  It is clear that 
the wavefunction of the full system is non-singular at the origin of
field space.  This does not agree with the asymptotics of \bswave\ ,
and we expect this disagreement to become important for $r < \gs^{1/3}$.

We can construct a system where the moduli space approximation is
accurate for arbitrarily small $r$ by considering
a 0-brane in the presence of $N$ 4-branes and letting $N \rightarrow
\infty$.   Introducing the large $N$ coupling $\tilde \gs =\gs N$
we find that the characteristic moduli space length scale is 
${\tilde \gs}^{1/3}$ while the characteristic velocity is
$\sim {\tilde \gs}^{2/3}/N$.  This low velocity shrinks the stadium
size where Born-Oppenheimer ceases to be valid down to $\sim
{\tilde \gs}^{1/3}/N^{1/2}$, arbitrarily smaller
than the throat size of the moduli
space.

\newsec{Non-trivial topology from open string theory}

By now we have seen many examples in which at very short distances
effects which are normally described by a non-trivial background
metric are instead described by the non-trivial physics of an open
string theory.

A different and very strong test of this alternate description is to
consider a background space of non-trivial topology.  The claim is now
that a space which at long distances must be described by introducing
several coordinate patches, can at short distances be described as the
moduli space of the gauge theory on the world-volume of D-branes in
flat space.

Consider a space with a non-contractible cycle.  If the volume of the
cycle is large compared to the string scale, the metric description is
appropriate.  D-branes moving on the space are well described by
simply inserting the metric into the world-volume action.  Even if
they approach to within the string scale, the associated open string
dynamics will take place within a topologically trivial region of
small curvature.

If the volume of the cycle is small
compared to the string scale, its existence need not
be manifest in the metric description, because the objects which see only
the metric (such as closed strings) are too large to unambiguously see it.
But, if space-time is a sensible concept on
scales shorter than the string scale,
the cycle must be visible in the D-brane gauge theory
moduli space.
Since we can go continuously from one
regime to the other, the same topology must be reproduced in the two
limits by the two descriptions.

An example of this was found in \dm.
As is well known, a $\BC^2/\BZ_n$ orbifold fixed point
singularity can be resolved to a smooth ALE manifold with
$H^2\cong \BZ^{n-1}$, and this region of string theory moduli space
is smooth.
The full moduli space of such compactifications was studied in
\Aspinwall, where it is verified that
the large volume and orbifold limits are connected.

In \dm, the blowup of this orbifold singularity was studied
through its effect on the world-volume gauge theory of a system of D-branes.
The moduli of the orbifold theory are in closed string twist sectors.
Turning them on changes couplings (Fayet-Iliopoulos terms)
in the open string theory, and changes the topology of its
moduli space from an orbifold to its resolution, an ALE space.

As we will review shortly, this theory is a linear sigma model,
which reduces at low energy
to a conventional non-linear sigma model with ALE target.
Such linear sigma models are very useful devices for studying
moduli spaces of string compactifications \WitPhase, as they allow
continuous transition between
models with different space-time topology and even models
(such as Landau-Ginzburg models)
with no direct space-time interpretation.
For this purpose, the precise UV definition of the model is not important;
since a perturbative string compactification is defined by specifying
a conformal field theory,
any model which flows to the same conformal fixed points in the
IR can be used.

If one considers processes at finite energy,
one can explore more of the configuration space of the linear sigma model.
This physics depends on the specific
UV definition of the model, and was not meaningful in the context of
perturbative fundamental string theory.

In the D-brane context, it is meaningful.  D-brane world-volume theories
have preferred scales, which can depend on the string coupling and the
other moduli.
Indeed, the super Yang-Mills actions we have been using are defined in
static gauge, for which world-volume and space-time scales are identified.
Thus finite energy D-brane scattering involves finite energy world-volume
physics.

Taking the low-energy limit of the linear sigma model involves
restricting to minima of the potential and quotienting by the action
of the gauge group.  At low but finite
energy, the system may not sit at a minimum,
but will travel close to minima, and the topology of the available region
of configuration space will be the same as that of the moduli space.

However, if the available energy is greater than the height
of a potential barrier, the available region
of configuration space can have a different topology.
This leads to the possibility that the
effective topology of the space-time could be different at finite
energy, a striking modification of our usual ideas.

We proceed to show that this is indeed true in two examples.

\subsec{D$0$ motion on an Eguchi-Hanson space}

We start by considering the
scattering of a D$0$ brane from the fixed point in the orbifold
$\BC^2/\BZ_2$.
A D-brane moving on this orbifold is described by starting with the
theory of a D-brane and its image on $\BC^2$, a $U(2)$ gauge theory.
One then projects on
states invariant under a $\BZ_2$ acting simultaneously on space-time
and on the D-brane (Chan-Paton) index, obtaining a $U(1)\times U(1)$
gauge theory, given in \refs{\dm,\PolNew} .
The $\BZ_2$ action can be taken to preserve the diagonal
$U(1)\times U(1)\subset U(2)$ gauge symmetry, and then
the surviving D-brane coordinates in $\BC^2$
are the off-diagonal components of the $2\times 2$ matrix,
$$
X^6 + iX^7 = \left(\matrix{0&X_{01}\cr X_{10}&0}\right)
\qquad
X^8 + iX^9 = \left(\matrix{0&\bar X_{01}\cr \bar X_{10}&0}\right).
$$

The physics is essentially that the $0$-brane will scatter off its
image, leading to results very similar to the D$0$-D$0$ scattering
of section 3.  Some quantitative results change, for example
half of the $16$ fermion zero modes of the stretched strings
in that problem are projected out, but the qualitative nature of the
scattering and in particular softness at the scale $l_P^{11}$ remain.

Eguchi-Hanson space is the ALE space
obtained by blowup of this orbifold.
As shown in \dm, the background moduli
$\zeta$ controlling the blowup couple as
Fayet-Iliopoulos terms in the open string gauge theory.
The $0$-brane moduli space is derived by restricting to minima of the potential
and quotienting by the action of the gauge symmetry.
The problem at hand has enough supersymmetry to determine the potential
uniquely from the gauge action, and this guarantees that
this will be a hyperk\"ahler quotient construction \HKLM,
indeed the one proposed for this purpose in \Kronheimer.

We quote the resulting potential (\dm, equations (6.21) and (8.6)):
\eqn\ztwopot{\eqalign{
V &= {1\over g_s}\left[|\mu^C - \zeta^C|^2 + (\mu^R - \zeta^R)^2\right] \cr
\mu^C &= X_{01} \bar X_{10} - \bar X_{01} X_{10} \cr
\mu^R &= |X_{01}|^2 - |X_{10}|^2 + |\bar X_{01}|^2 - |\bar X_{10}|^2.
}}
The moduli $\zeta^C$ and $\zeta^R$ become the periods
(integrals over the non-trivial
two-cycle) of the holomorphic two-form and K\"ahler form
respectively.

The D$0$-brane scattering calculation could now be repeated by first
finding the classical trajectory, and then computing the one-loop
phase shift \phase\ along this trajectory.
At finite energy, the trajectory will not be constrained to the minima
$V=0$ but can of course explore all $V<E_{kin}=v^2/2g_s$.

The introduction of non-zero $\zeta$ leads to a qualitatively new feature
in the potential: there is a (unique)
saddle point at $X=\bar X=0$, of height
$|\vec\zeta|^2 \equiv |\zeta^C|^2+(\zeta^R)^2$.
Since $\zeta$ is at our disposal, given any finite incoming energy $E_{kin}$,
there will be a geometry with $\zeta$ small enough so that motion over
this potential barrier will be completely unconstrained.
The moduli space approximation will break down dramatically, and in fact
(in this example) the available configuration space will have
trivial topology.

Turning this around, given an ALE space with periods $|\vec\zeta|<<1$,
so the two-cycle is sub-stringy,
we see that the available configuration space has ALE topology only if
$v^2 << |\vec\zeta|^2$.
A sufficiently
fast (but non-relativistic) D-brane, with
$$v^2 > |\vec\zeta|^2$$
and momentum of order $v/g_s \roughly< 1/R_{11}$,
will move not on the ALE space
but in a `stadium' of trivial topology.
This region is of size $\sim v^{1/2}$ and
using the criterion of section 3, we see that
the Born-Oppenheimer approximation breaks down
in the same region.

At very small $\zeta$, quantum effects will wash out the two-cycle at
any finite energy.  If we take $\zeta \sim (l_P^{11})^2 \propto
g_s^{2/3}$, the rescaling of \scaledvars\ can be applied to scale $g_s$
out of the action.  For $\zeta<< (l_P^{11})^2$, the quantum mechanics
will be dominated by the quartic $X^4$ potential, while the $\zeta
X^2$ corrections can be treated perturbatively.  Thus the physics is
qualitatively the same as in the orbifold limit, and no vestige of
non-trivial topology remains.  In particular, possible bound states or
resonances in this scattering will again be associated with the scale
$l_P^{11}$.
It will be important to understand the relationship between
the small sizes discussed here and those discussed in \Aspinwallmin.

\subsec{Small instantons}

Another example is a gauge bundle with non-trivial topology.  Let us
consider supersymmetric backgrounds of the type \I\ string on
$\BR^6\times \BR^4$, with self-dual gauge field, constant in $\BR^6$.
Such configurations are classified topologically by the second Chern
class $c_2 = {1\over 8\pi^2}\int \tr F^2$.

As Witten argued \witten, such a configuration is a point in the
moduli space of a system of $c_2$ D$5$-branes, and thus a self-dual
gauge field is described either by a non-trivial $9$-brane background
or by a $59$-brane system.  This can be seen more explicitly by
studying a D$1$-brane probe \dgauge.  Its basic world-sheet theory is
that of the heterotic string, but in the presence of a $5$-brane, its
world-volume theory has additional fermion zero modes (1-5 strings),
and position-dependent Yukawa couplings determined by the expectation
values of $5$-brane fields (5-9 strings).  Integrating out massive
fermions corresponds to projecting on the massless subspace, which
induces a non-trivial gauge connection, the self-dual gauge field.

Although both the original gauge field in the $9$-branes and the
equivalent 5-brane configuration are open string field theory
configurations, the equivalence between the two has the same character
as the open-closed string problems we have been discussing.  On the
one hand, the 9-brane gauge field could be seen (at leading order) in
the o.p.e. of two 5-9 vertex operators, and it would then couple to a
bilinear in the D$1$-brane gauge fermions.  The cross channel of this
four-point diagram involves the o.p.e.  of the 5-9 and 1-9 vertex
operators, producing the intermediate 1-5 strings of the D-probe
construction.

The analog of the phenomenon we saw for the Eguchi-Hanson space is
that, at energies larger than the instanton scale size, the effective
theory of the massless modes is inadequate, and the D$1$-brane couples
to a larger rank gauge bundle of trivial topology.  In this sense, a
sufficiently small instanton is `washed out' at finite energy.

{}From \dgauge, the scale of the fermion Yukawa couplings is equal to
the instanton scale size $\rho$, and thus the condition for this to
happen is simply $E > \rho/\ap$.

\newsec{Bound states of 1-branes}

Type \IIb\ string theory has a family of soliton strings.  The
solitons are labeled by two relatively prime integers $(m,n)$ which
specify their couplings to the (RR, NS-NS) two-form potentials
\Schwarz.  The $(m,n)$ soliton string can be constructed as a bound
state of $m$ Dirichlet 1-branes and $n$ fundamental strings
\WittenBound.  In this section, we study the dynamics of the $(2,1)$
soliton string, and show that the size of the bound state is given by,
up to logarithms, $\gs^{1/2} \ls$.

The physical significance of this length scale is unclear.  Perhaps it
has an origin in F-theory?  It is the tension scale of the 1-brane,
$T_{\rm 1-brane}^{-1/2} = \gs^{1/2} \ls$.

The upshot of the analysis given in section 7.2 is that the world-sheet
action for the $(2,1)$ soliton string includes eight scalar fields
$\phi_i$ which act as relative coordinates for the two 1-branes.  With
$\phi_i$ normalized to measure the separation in string units, the
world-sheet effective action is
$$
\int d^2 x \,\, {1 \over 2 \gs} (\partial \phi_i)^2
- V(\phi) \,.
$$
A $1/\gs$ appears in front of the kinetic term, reflecting its origin
in an open string disc diagram.  The one-loop effective potential
$V(\phi)$, obtained in section 7.2, is shown in Fig.~1.  The exact
expression for the effective potential is somewhat complicated, but
roughly speaking the one-loop potential is attractive at separations
less than $\gs^{1/2} \ls$.  In this range the fields $\phi_i$ are
massive, with a mass $m_\phi$ of order $\gs^{1/2} \ms$.  At
separations greater than $\gs^{1/2} \ls$, the effective potential
flattens out, and the fields $\phi_i$ become massless.

\ifig\figone{Effective potential (in units of $\gs \ms$) vs.
1-brane separation (in units of $\gs^{1/2} \ls$).}
{
\epsfbox{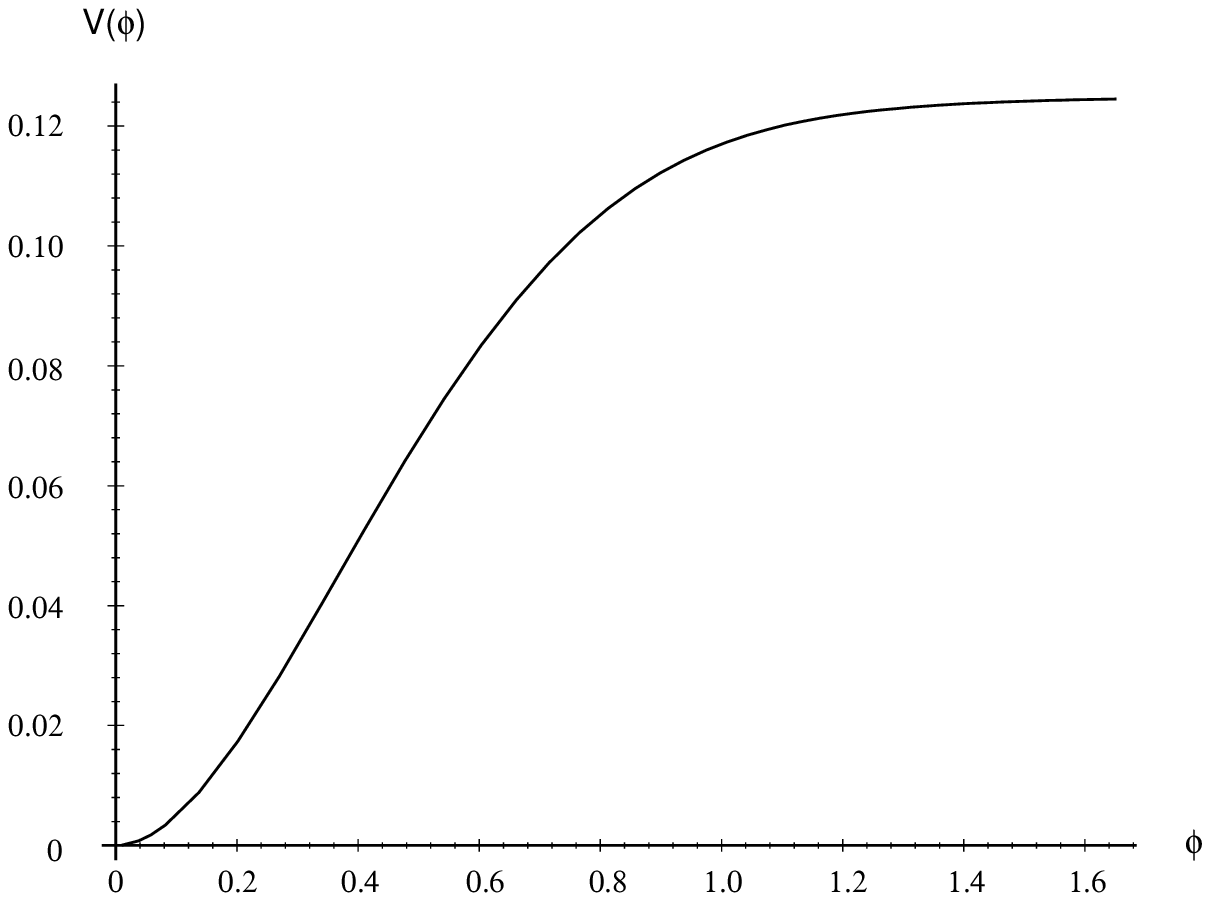}}

Assuming that the  1-branes are localized near the minimum of $V$,
we can calculate the mean square size of the
$(2,1)$ bound state in free field theory.
\eqn\size{\eqalign{
<{\scriptstyle \sum}_i \phi_i^2> &= 8 \gs \int^\Lambda {d^2k \over
(2\pi)^2} \, {1 \over k^2 + m_\phi^2}\cr
&\sim \gs \log {\Lambda \over m_\phi} \,. \cr}
}
The ultraviolet cutoff $\Lambda$ has the interpretation that we choose
to measure the size of the bound state averaged over a region of size
$\Lambda^{-1}$.  In our analysis we neglect massive string modes, so
$\Lambda$ should be taken to be somewhat below the string scale.  Note
that the mean separation between the 1-branes is, up to logarithms,
$\gs^{1/2} \ls$.

The size of the $(2,1)$ bound state may receive loop corrections,
which shift the coefficient of $\gs^{1/2} \ls$.  We discuss this
further in section 7.3, where we also consider certain resonances
above the $(2,1)$ bound state, for which loop corrections are
controllably small.

\subsec{Finite radius effects and T-duality}

It is possible to estimate the size of a $(2,1)$ bound state that
has been wrapped on a circle, a configuration which is T-dual to a
bound state of two 0-branes.  This is of interest for two reasons.  It
shows that our understanding of 0-brane and 1-brane dynamics is
consistent, and it also illustrates that T-duality is compatible with
non-trivial dynamics occurring at distances shorter than the string
scale.

Consider a $(2,1)$ soliton string wrapped on a circle of radius $R$.
When $R$ is sufficiently small, the size of the bound state can be
found by scaling analysis, as follows.  We will formulate the
world-sheet action for the $(2,1)$ soliton string in more detail in
section 7.2, but let us suppose that the circle is small enough that
we can use dimensional reduction to get an effective action
just for the zero modes.  The resulting effective action is identical
to the action for 0-branes, but with an effective coupling $g_{\rm
eff} = \gs \ls / R$.  A scaling analysis of the 0-brane action
\refs{\DanFerSun, \kp}, 
reviewed in section 3, shows that the only length scale in the problem is
$g_{\rm eff}^{1/3}
\ls$.  We therefore expect the size of the $(2,1$) bound state to be
set by $g_{\rm eff}^{1/3} \ls = (\gs \ls / R)^{1/3} \ls$.

When is this dimensional reduction valid?  We expect it to break down
as $R$ becomes large.  Compare the energy of a Kaluza-Klein
excitation, $E_{KK} \sim 1/R$, with the energy scale of the effective
0-brane problem.  That energy is given by scaling analysis as $E_0 =
g_{\rm eff}^{1/3} \ms = (\gs \ls /R)^{1/3} \ms$ .  The dimensional
reduction is only valid when $R < \gs^{-1/2} \ls$, so that the
Kaluza-Klein modes are frozen out.

When $R > \gs^{-1/2} \ls$ the scaling analysis breaks down.
Fortunately, in this regime the size of the $(2,1)$ bound state is
accurately given by the one-loop calculation \size.  To justify using
the one-loop calculation, note that in this regime the mass gap
$m_\phi \sim \gs^{1/2} \ms$ is large compared to the spacing between
energy levels $\sim 1/R$.  So finite size effects on the world-sheet of
the $(2,1)$ soliton string are unimportant.  The size of the bound
state is the same as if $R$ was infinite, namely $\gs^{1/2} \ls$.

To summarize, the size of the $(2,1)$ bound state is given by
$(\gs \ls /R)^{1/3} \ls$ up to $R \sim \gs^{-1/2} \ls$, and above that radius
the size $\sim \gs^{1/2} \ls$ is independent of $R$.  Let us now
re-express this in \IIa\ string theory, with T-dual radius and
coupling
$$
R_A \sim \ls^2 / R \qquad g_A \sim \gs \ls / R \,.
$$
Under T-duality, the $(2,1)$ soliton string becomes a bound state at
threshold of two 0-branes with one unit of Kaluza-Klein momentum, as
in \SenBound.  Re-expressing the size of the $(2,1)$ bound state in
\IIa\ language, we expect that the size of the 0-brane bound state at
threshold should be $g_A^{1/3}\ls$ when $R_A > g_A^{1/3} \ls$, while for
$R_A < g_A^{1/3} \ls$ the size should be $(g_A \ls /R_A)^{1/2} \ls$.

This meshes with our understanding of 0-brane dynamics.  Again,  scaling
analysis shows that the size of the 0-brane bound state at threshold
in flat space is $g_A^{1/3} \ls$.  This should not change upon
compactification, provided the compactification scale $R_A > g_A^{1/3}
\ls$.

The $(g_A \ls /R_A)^{1/2} \ls$ scale can also be understood, as follows.  The
two $0$-branes in the bound state at threshold have one unit of
Kaluza-Klein momentum.  Semiclassically, this corresponds to a
relative velocity $v \sim g_A \ls/R_A$.  As in our discussion of 0-brane
scattering, the $0$-branes interact in a stadium of size $\sim v^{1/2} \ls
\sim (g_A \ls/R_A)^{1/2} \ls$.  We identify the size of the stadium with the
size of the bound state.

\subsec{One-loop effective potential}

To formulate the world-sheet dynamics of the $(2,1)$ string we follow
\WittenBound.  The degrees of freedom relevant for low-energy
dynamics are the massless modes of the open strings which end on the
branes.  That is, the relevant world-sheet degrees of freedom of the
$(2,1)$ string are
$$
\eqalign{
A_\alpha &= {i \over 2} \left(A_\alpha^0 \identity + A_\alpha^a
\sigma^a\right)\cr
\phi_i &= {i \over 2} \left(\phi_i^0 \identity +
       \phi_i^a \sigma^a\right)\cr
\psi &= {i \over 2} \left(\psi^0 \identity +
       \psi^a \sigma^a\right)\cr}
$$
where $a$ is an adjoint $SU(2)$ index, $A_\alpha$ ($\alpha = 0,1$) is
a 1+1 dimensional $U(2)$ gauge field, $\phi_i$ ($i=2,\ldots,9$) is a
collection of eight adjoint Higgs fields, and $\psi$ is a sixteen
component adjoint spinor.  These degrees of freedom are actually
sufficient to describe the dynamics of two D-1-branes and an arbitrary
number of fundamental strings, because the presence of fundamental
strings corresponds to turning on a background $U(2)$ gauge field
\WittenBound.  The configuration of interest, with a single
fundamental string present, corresponds to the background field
produced by placing a charge at infinity in the fundamental of $U(2)$.

For a world-sheet action we take the dimensional reduction of $\CN=1, D=10$
supersymmetric Yang-Mills theory.  The action comes in two
decoupled parts.  One part, involving the $U(1)$ degrees of freedom,
governs the center of mass motion of the system.
$$ S_{\rm cm} = \int d^2x \, - {1 \over 4 \gs}
\left(F_{\alpha\beta}^0\right)^2 - {1 \over 2 \gs}
\left(\partial_\alpha \phi_i^0\right)^2 + {i \over 2} \bar\psi^0
\Gamma^\alpha \partial_\alpha \psi^0
$$
This action has the right form and correct massless degrees of freedom
to be the world-sheet action of a $(2,1)$ soliton string
\refs{\CalKle,\Schmid} .  A second part, which involves the $SU(2)$
degrees of freedom, controls the relative motion of the 1-branes.
$$\eqalign{
S_{\rm rel} = \int d^2x \, - {1 \over 4 \gs} &\left(F_{\alpha\beta}^a\right)^2
- {1 \over 2 \gs} \left({\cal D}_\alpha \phi_i^a\right)^2 - {1 \over 4 \gs}
\vert \phi_i \times \phi_j \vert^2  \cr
&+ {i \over 2} \bar\psi^a \Gamma^\alpha
{\cal D}_\alpha \psi^a + {i \over 2} \epsilon^{abc} \phi_i^a \bar\psi^b
\Gamma^i \psi^c
}$$
In the presence of the background field corresponding to a single
fundamental string, it has been shown that this $SU(2)$ theory
generates a mass gap above a single supersymmetric vacuum state, so
that the relative motion of the 1-branes does not contribute to the
low-energy dynamics of the soliton string \WittenBound.

We proceed to analyze the dynamics of the relative $SU(2)$ theory in
more detail.  We wrap the two D-1-branes around a circle of
radius $R$ in the $x^1$ direction.  Classically, the theory has
a moduli space of vacua, parameterized by the Higgs fields which
break $SU(2)$ to $U(1)$.  Up to gauge and global symmetries, such a
Higgs field takes the form
\eqn\Higgs{
< \phi_2^a > = (0,0,b)
}
with the interpretation that the two D-1-branes are separated a
distance $b$ in the $x^2$ direction.  To include fundamental strings,
we turn on a background electric field in the unbroken $U(1) \subset
SU(2)$, by giving a time-dependent expectation value
\eqn\gauge{
< A_1^a > = (0,0,vt) \,.
}
Classically, this electric field has an energy density ${1 \over 2
\gs} v^2$, and supersymmetry is broken for $v \not= 0$.  When $b$ is
large, $SU(2)$ is broken to $U(1)$ at high energies, and the classical
calculation should be accurate.  This shows that there is an energy
barrier to making $b$ large \WittenBound.  For supersymmetry to be
unbroken at $b=0$, the energy of the background electric field
must somehow be cancelled.

We now show that this cancellation occurs at one loop, via the
generation of an effective $\theta$-angle.  Integrating out the
massive $SU(2)$ degrees of freedom in the background \Higgs, \gauge~
produces an effective action for the light $U(1)$ degrees of freedom.
The resulting one-loop determinants can be written in a proper-time
representation.
$$
S_{\rm one-loop} = - \int d^2x \, {v \over 8 \pi} \int_0^\infty
{ds \over s} e^{-s b^2} {1 \over \sin sv} \left(16 \cos sv - 4 \cos 2sv
- 12 \right)
$$
Note that $S_{\rm one-loop} = \int d^2x {v \over 2 \pi} \delta(b,v)$,
where $\delta(b,v)$ is the phase shift for 0-brane scattering \phase.

The one-loop effective action has an imaginary part, arising from the
poles in the integrand, which measures the rate for producing a pair
of unexcited open strings stretched between the two 1-branes.
$$
{\it Im} \, S_{\rm one-loop} = - \int d^2 x \, {2 v \over \pi} \log \tanh
{\pi b^2 \over 2 v}
$$
The pair production rate vanishes at large $b$ but diverges
logarithmically as $b \rightarrow 0$.  This reflects the fact for
small $b$ the strings do not want to stay in the unbroken $U(1)$
subgroup, but rather start to oscillate in all the non-commuting
$SU(2)$ directions.  In order to recover the space-time interpretation
from the $SU(2)$ dynamics, we must project the oscillations into a
$U(1)$ subgroup anyways, so we will simply drop the imaginary part of
the one-loop effective action from now on.  Moreover, as we will see
below, in the $(2,1)$ bound state $v$ is order $\gs$, so the pair
production rate can be made arbitrarily small by going to weak
coupling.

The real part of the one-loop effective action is finite.  At large
$b^2$, it behaves as
$$
{\it Re} S_{\rm one-loop} =\int d^2x \, {1 \over 2\pi}
{v^4 \over b^6} +{\cal O}({v^6\over b^{10}}) \, .
$$
One expects this $v^4 / b^6$ interaction at large separations to arise
from massless closed string exchange between the two D-1-branes.  We
see here another manifestation of the remarkable fact that a loop of
massless open strings can mimic the leading small velocity behavior of
massless closed string exchange.  At small $b^2$, the real part of the
one-loop effective action is
$$
{\it Re} S_{\rm one-loop} = \int d^2 x \, \half v - { 1 \over \pi}
(4 \log 2 - 1) b^2 + {\cal O}(b^6 / v^2) \,.
$$
Note that a term linear in the velocity, a $\theta$-term in the 1+1
dimensional gauge theory, has appeared in the effective action, as
required by the chiral anomaly.\foot{It seems difficult to
write a manifestly supersymmetric non-Abelian generalization of this
term \WittenGrass.}  Moreover, the order $b^2$ term in the expansion
is a mass term for the Higgs fields.

In 1+1 dimensions, a $\theta$-angle produces a shift in the background
electric field.  We will see that the coefficient is just what is
needed to produce a supersymmetric vacuum at $b=0$: it will cancel the
electric field which corresponds to the presence of a single
fundamental string.  The effective action for the unbroken $U(1)$
gauge field is
$$\eqalign{
S_{\rm eff}\left[A_1,\phi_i\right] &= \int d^2x \,
{1 \over 2 \gs} \dot{A}_1^2+ {\rm Re} S_{\rm one-loop} \cr
& \!\!\! = \int d^2x \,
{1 \over 2 \gs} \dot{A}_1^2 + \half \dot{A}_1 \cr
&- {i \over 2 \pi} \dot{A}_1 \log
\left\lbrace {\dot{A}_1 + i \phi^2 \over \dot{A}_1 - i \phi^2}
\left[{\Gamma\left(- {i \phi^2 \over 2 \dot{A}_1}\right)
       \Gamma\left(\half + {i \phi^2 \over 2 \dot{A}_1}\right) \over
\Gamma\left({i \phi^2 \over 2 \dot{A}_1}\right)
       \Gamma\left(\half - {i \phi^2 \over 2 \dot{A}_1}\right)}
\right]^4 \right\rbrace \,.\cr}
$$
plus the fermion terms required by supersymmetry.  We have dropped the
imaginary part of the one-loop action, and adopted Coulomb gauge, $A_0
= 0$, $\partial_1 A_1 = 0$.  The D-branes are wrapped on a circle of
radius $R$, which makes $A_1$ a periodic variable with period $2 / R$.
The momentum conjugate to $A_1$,
$$\eqalign{
{\cal E} &= {\partial L_{\rm eff} \over \dot{A}_1} \cr
         &= 2 \pi R\left({1 \over \gs} \dot{A}_1 + \half
+ \cdots \right)\cr}
$$
has a discrete spectrum, quantized in integer multiples of $\pi R$.
The background field corresponding to a single fundamental string,
${\cal E} = \pi R$, exactly cancels against the contribution of the
$\theta$-angle to the definition of the electric
field.\foot{The $\theta$-angle is not renormalized beyond one
loop, so this cancellation will persist to all orders in perturbation
theory.}

The effective potential for the Higgs fields is then
$$
V(\phi) = {1 \over 2 \pi R} \left.\left({\cal E} \dot{A}_1 - L_{\rm eff}
\left(A_1,\phi\right)\right)\right\vert_{\dot{A}_1 = \dot{A}_1(\phi)}
$$
where $\dot{A}_1(\phi)$ is determined by solving ${\cal E} = \pi R$.
Note that by rescaling the fields
$$
A_1 = \gs \tilde{A}_1 \qquad \phi = \gs^{1/2} \tilde{\phi}
$$
all dependence on the coupling constant can be eliminated, except for an
overall factor of $\gs$ multiplying the effective potential.
This shows that the effective potential only depends on the separation
measured in units of $\gs^{1/2} \ls$, and that the `velocity'
$\dot{A}_1$ corresponding to a single fundamental string is order
$\gs$, in accord with the T-duality considerations of the previous section.
The resulting potential, shown in Fig.~1, has the behavior
$$
V(\phi) \sim \cases{
{1 \over \pi} (4 \log 2 - 1) \left(\phi_i\right)^2 & as
$\phi \rightarrow 0$ \cr
\noalign{\vskip 4 pt}
{1 \over 8} \gs & as $\phi \rightarrow \infty .$ \cr}
$$
Supersymmetry is restored at the origin, where the effective potential
vanishes, and a mass gap $m_\phi = \gs^{1/2} \ms \sqrt{{2 \over \pi}
(4 \log 2 - 1)}$ is generated at one loop.

\subsec{Loop corrections to the bound state size}

Previously
we found the size of the $(2,1)$ bound state using the effective action
$$
\int d^2 x \, {1 \over 2 \gs} \left(\partial_\alpha
\phi_i\right)^2 - \gs V(\phi/\gs^{1/2}) \,,
$$
where $V(\phi)$ is the one-loop effective potential, obtained by
integrating out the Higgsed $SU(2)$ degrees of freedom.  Here we have made
all dependence on the coupling constant explicit.  The mean square
size of the bound state is given by the two-point function
$\langle~\left(\phi_i\right)^2~\rangle$. 
Approximating $\phi$ as a free massive
field gives the estimate $<\left(\phi_i\right)^2> \sim \gs \log
{\Lambda \over m_\phi}$, where $m_\phi \sim \gs^{1/2} \ms$.

We now discuss corrections to this free-field approximation, from
loops of the light Abelian fields $\phi_i$.  To estimate loop effects,
it is convenient to rescale $\phi = \gs^{1/2} \tilde{\phi}$.  This
gives $\tilde{\phi}$ a canonical kinetic term, with action
$$
\int d^2 x \, {1 \over 2 } \left(\partial_\alpha
\tilde{\phi}_i\right)^2 - \gs V(\tilde{\phi}) \,.
$$
On the world-sheet of the 1-brane, $\tilde{\phi}$ is dimensionless,
while $\gs$ has units of $({\rm mass})^2$.  Thus the $L$ loop
contribution to $<\tilde{\phi}^2>$, naively ${\cal O}\left(\gs^L
\right)$, must really be ${\cal O}\left({\gs^L / (\hbox{\rm mass
scale})^{2L}}\right)$.  In this infrared divergent 1+1 dimensional
field theory, the mass scale is provided by the infrared cutoff
$m_\phi \sim \gs^{1/2}$.  It follows that $L$-loop graphs are really
${\cal O}(\gs^0)$, and are not suppressed by making the coupling
small.

What does this mean for the size of the $(2,1)$ bound state?  Just on
these dimensional grounds we know that $<\tilde{\phi}^2> = {\cal
O}(\gs^0)$, that is, that $<\left(\phi_i\right)^2> = {\cal O}(\gs)$.
So the size of the $(2,1)$ bound state is indeed set by $\gs^{1/2}
\ls$, in accord with the free-field approximation, although with a
numerical coefficient that can only be calculated by re-summing the
perturbation series.

We conclude by discussing a system in which loop corrections to the
size of the bound state can be made controllably small.  Consider a
resonant state in which $n$ units of background electric field are
present in the relative $SU(2)$ theory.  Such a state is unstable, and
will decay back down to the $(2,1)$ ground state (for $n$ odd), but
the decay rate can be made arbitrarily small by going to weak string
coupling.

We now show that the size of this resonance can be calculated in an
expansion in $1/n$.  This follows from the form of the effective
potential, which we determine as in the previous section.  The
background electric field ${\cal E} = n \pi R$ implies $\dot{A}_1
\approx \gs n / 2$, up to corrections which are order $n^0$.  The
effective potential is then
$$\eqalign{
V(\phi) &= {1 \over 2 \pi R} {\cal E} \dot{A}_1 - {1 \over 2 \gs} \dot{A}_1^2
- {1 \over 2} \dot{A}_1 + \dot{A}_1 f\left({\phi^2 \over \dot{A}_1}\right)\cr
&\approx {\rm const.} + {\gs n \over 2} f\left({2 \phi^2 \over \gs n}\right)
\cr}
$$
where
$$
f(x) = {1 \over \pi} (4 \log 2 - 1) x + {\cal O}\left(x^3\right)\,.
$$
The first term in the expansion of $f$ is a mass term for $\phi$.
Note that the mass is independent of $n$, that is, the infrared cutoff
in the theory does not depend on $n$.  But the interaction terms in
the effective potential are suppressed by powers of $1/n$, so
perturbation theory is an expansion in $1/n$.  At leading order, the
size of the resonance is given by the same free-field expression as
for the true $(2,1)$ bound state, $\gs^{1/2} \ls$, but with
corrections which are suppressed by powers of $n$.

\newsec{Effective field theory of open and closed strings}

In section 2, we showed that at leading order in the string coupling
constant, the long-distance interaction between D-branes was described
by supergravity, while the short-distance interaction was described by
the D-brane world-volume theory of the lightest open strings.

We now argue that this is true to all orders in the string coupling
constant.  As we saw in the context of the annulus, the essential
point is to be able to divide the moduli space of string world-sheets
into open string IR and closed string IR regions.  Singularities of
the short-distance interaction are controlled by the open string IR
region, while the leading long-distance behavior is controlled by the
closed string IR region.

Thus we need a decomposition of each higher genus moduli space
integral generalizing \opdecomp, in which each region can only
contribute to open string IR or closed string IR singularities.  We
will then be able to adapt the argument of subsection 2.3 to bound the
contribution of the `wrong' channel in each region.

This can be done by associating points in moduli space with diagrams
built by attaching open string and closed string propagators to
vertices, as in string field theory.  The IR region is just the long
proper-time region for that propagator, while the condition that a
propagator not contribute to singularities in the wrong channel can be
fulfilled by a cutoff eliminating short proper time.  As in section 2,
the space-time effect will be to make the closed string propagator soft
at sub-stringy scales, and cut off the open string theory at string
scale in the UV.

A higher order diagram can contain many propagators, of course, and a
given region in moduli space might contain both open and closed string
IR regions, for different propagators.  The general statement is that
every singularity of an amplitude can be reproduced by a low-energy
effective field theory containing three types of fields: the
supergravity fields from massless closed strings,
open string fields which are massless everywhere in
configuration space (such as the coordinates of a single D-brane),
and open string fields which can become massless at special points of
configuration space (from strings stretched between branes).
All other massive string modes can be dropped.  Each region of
string moduli space will then correspond
 to a particular diagram of
this effective field theory.

Such a decomposition of moduli space was developed by Zwiebach
\zwiebach.  The strategy is to associate to each point in complex
moduli space a minimal area world-sheet metric satisfying certain
conditions.  For the decomposition of \zwiebach, the conditions are
that topologically non-trivial closed curves (describing possible
curves on which factorization on closed strings can be done) have
length greater or equal to $2\pi$, and non-trivial open curves (which
cannot be contracted while keeping each end on a boundary) have length
greater or equal to $\pi$.

Such a minimal area metric will be a connected sum of components which
can each be identified as a vertex or propagator with specified
moduli.  A propagator is a cylinder or strip with proper time
integrated from zero to infinity.  The vertex components will not be
the overlap-type vertices of Witten's string field theory
\WitSFT\ but are defined to include `stubs' and `strips,' definite length
closed and open string propagators.  Combining these with the propagators
implements the lower bound on the proper time in each channel, and thus
this decomposition fulfills our requirements.

The basic vertices are the following, taken from \zwiebach\ section
3.1 but listed in a different order.  First, there is a three open
string vertex, for which the world-sheet is the Witten vertex with
strips of length $\pi$ attached to the legs.  This will include
interactions on a D-brane world-volume.  Second (iv in \zwiebach),
there is a three closed string vertex, whose world-sheet is the double
of the preceding one.  This includes bulk interactions.  Third (iii in
\zwiebach), there is a disk with a single closed string puncture,
representing a closed string ending on a D-brane.  In particular, the
basic sources of metric, RR field and dilaton are represented by this
diagram.  Finally (ii in \zwiebach), there is a vertex coupling one
open and one closed string, with a stub for each.  This will produce
open-closed couplings of the sort computed in
\compsource\ and described in \refs{\douglas,\ghm}.

These are conveniently represented by diagrams which use double lines
to represent open string propagators and wavy lines to represent
closed string propagators.  All Feynman diagrams constructed from the
four basic vertices in figure 2 will appear.  In an action with $1/g$
and $1/g^2$ normalized kinetic terms, their coupling dependences are
$1/g$, $1/g^2$, $1/g$ and $1/g$, while if we rescale the fields to get
canonically normalized kinetic terms, their coupling dependences are
$g^{1/2}$, $g$, $g^0$ and $g^{1/2}$.  Higher point vertices can also
appear -- a $(k+2)$-string vertex acts like a tree diagram built from
$k$ $3$-string vertices in this counting.

\ifig\figtwo{Vertices of open-closed string effective theory.}
{\epsfxsize4.0in\epsfbox{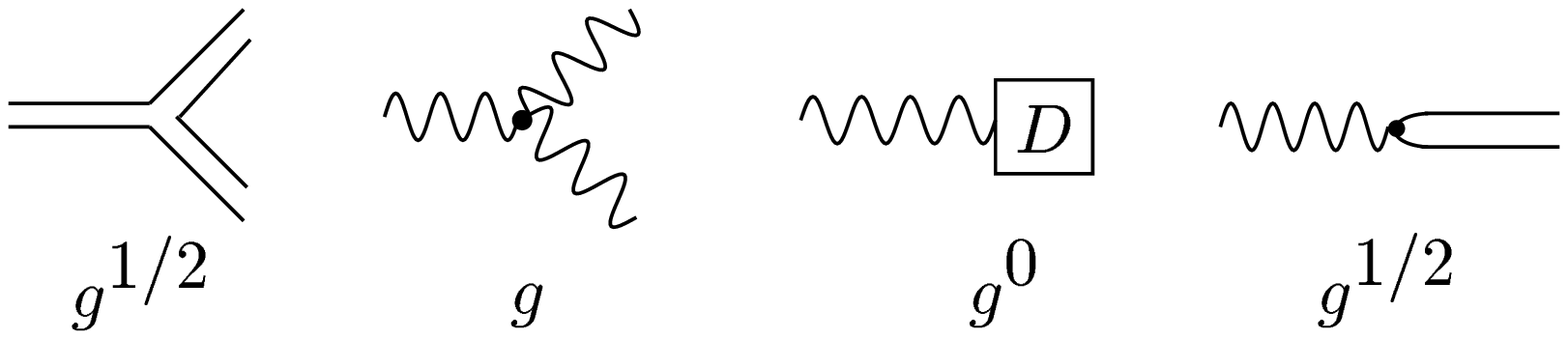}}

The basic statement is then that the UV region (short proper time) in
any open string loop or closed string propagator is replaced by a long
proper-time limit in another diagram.  The simplest example is the
annulus, which now becomes a sum of two diagrams, the open string loop
and the closed string exchange.

The next order is at $O(g)$ and the contributions include two-loop
open string diagrams, an interaction between three closed strings
emitted by branes, and mixed open-closed diagrams, all given in figure
3.

\ifig\figthree{Diagrams at $O(g)$.}
{\epsfxsize5.0in\epsfbox{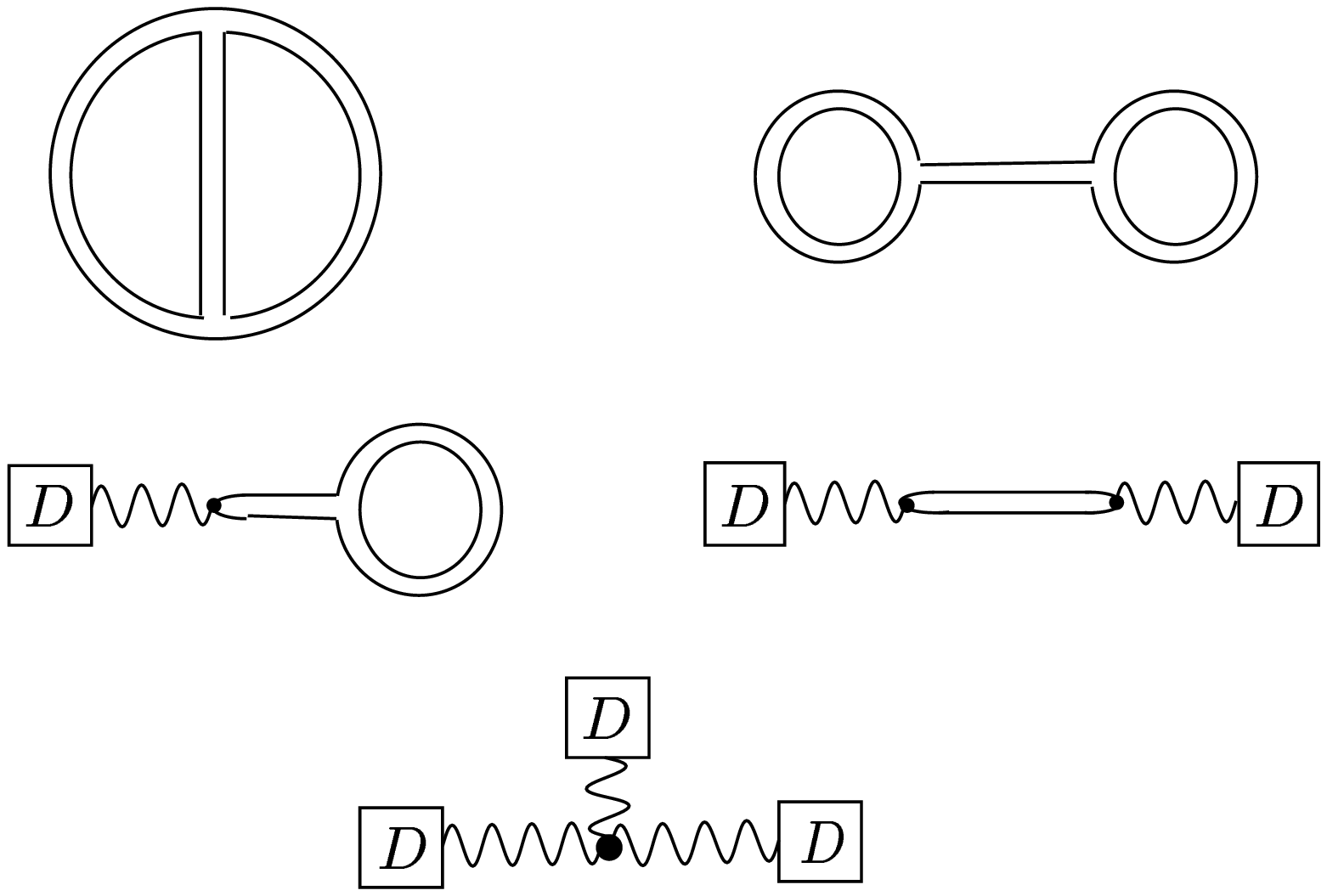}}

There are three world-sheet boundaries to associate with branes, and
each diagram would appear with all possible labellings.  As one
follows figure 3 downwards, one can see how each open string loop, in
the short proper-time limit, is replaced by a closed string
propagator.

We believe that this qualitative discussion suffices to demonstrate
the principles we used in this work.  While it might be possible to
systematically derive a low-energy effective Lagrangian from string
field theory, in this work we implicitly adopted a hybrid approach,
using open and closed string propagators as defined here, but
computing the vertices for interactions of massless fields using other
approaches.  In particular, the higher point vertices were inferred
using considerations of gauge symmetry and supersymmetry.

We also believe that a stronger statement can be made: there exists an
effective field theory of D-brane world-volume field theories coupled
to supergravity, which reproduces the physics of non-relativistic
D-branes and the closed string fields they produce, at leading order
in an $\ap$ expansion.  The development of such a formalism would be a
valuable aid to further research.

Let us also mention that in general D-brane world-volume theories are
not finite as quantum field theories.  Indeed, there are even
non-renormalizable examples, such as a $4$-brane in the background of
an $8$-brane.  In the present work we did not need to discuss this,
because we only considered $0$-brane dynamics in detail.

Such open string UV divergences should be reinterpreted as closed
string IR divergences.  For example, in the D4-D8 example, the growth of
the $4$-brane gauge coupling with energy will be re-interpreted as the
linear growth of the $8$-brane field with distance
\refs{\PolWit,\eightb}.
A complete effective field theory would come with
a specific open string UV cutoff,
matched with a closed string propagator modified to become non-singular
at sub-stringy distances, such that the sum was independent of the choice
of cutoff.  The two elements would correspond to the two terms in
\opdecomp.

\newsec{Conclusion}

Let us summarize our findings.  We argued that D-brane dynamics
changes character at length scales $r \sim l_s$.  At distances much
larger than $l_s$ massless closed string modes, gravitons, dilatons,
and gauge fields, mediate interactions.  This is the domain of
supergravity.  But, perhaps surprisingly, notions of distance continue
to make sense for $r < l_s$.  Closed string effects are soft at such
short distances and the dominant interactions are mediated by the
lightest open string states stretched between D-branes.  The dynamics
of these lightest open string states and hence the short-distance
behavior of D-branes is described by a quantum field theory on the
D-brane world-volume.

After establishing that D-brane quantum theory was the correct tool
for studying the sub-stringy domain we proceeded to explore it.  We
emphasized the use of D$0$-branes and the associated world-line quantum
mechanics.  To explore the structure of the 0-branes themselves we
scattered them off each other, extending the work of
\Bachas.  We found evidence for length scales of $\gs^{1/3} l_s
\sim l_P^{11}$ and $\gs l_s \sim R_{11}$.

{}From a conceptual point of view the occurrence of these sub-stringy
scales flows from a combination of the lack of scale invariance of
the world-brane quantum theory and the small space-time distance -- 
world-brane IR connection.  For 0-branes the world-line action can be
written schematically as
\eqn\wl{\int dt~ {\dot \phi}^2 + \gs \phi^4 ~.}
This is a non scale invariant theory as $\gs$ has world-line
dimensions of $({\rm mass})^3.$ So there will be characteristic
phenomena at world-line mass scales $\sim \gs^{1/3}$.  But for the
lightest stretched string $m \sim r$ so $r \sim \gs^{1/3}$ will be a
characteristic space-time scale.

$R_{11}$ appears in the size of higher derivative terms in \wl.  The
fine structure in the bound state and resonance spectrum of such
systems is governed, as in atomic physics, by the Compton wavelength
of the constituents.  For 0-branes this wavelength is $1/m_0 \sim g_s
l_s \sim R_{11}$ and so the fine structure provides indirect evidence
for this scale, as does the fine spacing of resonances at high energy.

At momentum transfers $\sim 1/R_{11}$ the zero-brane becomes
relativistic and the quantum mechanics approximation breaks down.  It
is striking that the IR world-volume approximation remains valid when
the 0-brane energies and momenta are far greater than $m_s$.
Excitation of massive open string modes is velocity-suppressed and can
be ignored for $v << 1$ independent of $\gs$.  Asymptotic
supersymmetry only becomes badly disturbed at $v \sim 1$.  This
suggests that D$0$-branes are small because they can have enormous
momentum $\sim m_s/\gs$ before they become relativistic and excite the
infinite tower of string excitations that comprise the `stringy halo'.
{}From this point of view fundamental strings are of string size
because they become relativistic at momenta $\sim m_s$ and excite
their stringy halo.  Could it be the degree of asymptotic
supersymmetry that determines the degree of locality?  A possible link
between these two basic notions is an intriguing prospect.

Armed with strong evidence that our probes were sharp, we used them to
explore sub-stringy geometry.  Our first example was the background
geometry around a 4-brane, which we studied using a 0-brane probe.  We
calculated the metric on the 0-brane moduli space.  We then looked for
the bound state at threshold predicted from M-theory considerations,
and found it in the moduli space approximation as a zero mode of a
Laplacian on the moduli space.  The bound state had size $\gs^{1/3}l_s
\sim l_P^{11}$ because this scale appeared in the moduli space
metric, as expected from scaling arguments
\refs{\DanFerSun,\kp}.  We should stress that this behavior is very
different from that of bound states of traditional `fat' solitons
like BPS monopoles \refs{\GibMan,\Senbdmon}.  Bound states of such
objects have size $1/m_W$, their classical size, independent of the
coupling $g^2$.  This is reflected in the metric on the two-monopole
moduli space, the Atiyah-Hitchin metric, which has features of
characteristic size $1/m_W$ and no $g^2$ dependence.

This example gives a particularly striking example of the difference
between the moduli space approximation and the finite energy dynamics.
The point $r=0$ where the 0-brane touches the 4-brane is at infinite
distance in the moduli space metric, but at finite distance in the
classical 0-brane action.  When the Born-Oppenheimer approximation
breaks down, the infinite distance found in the moduli space metric
becomes meaningless.  No single geometrical picture describes all the
dynamics.

Building on the work of
\refs{\dm,\dgauge,\PolNew} which studied the motion of D-branes
on spaces with blown up orbifold singularities and small instantons,
we studied the regime of validity of the geometrical description found
there.  In the orbifold example, at low velocity it is valid from
sub-stringy $S^2$ volume all the way down to volumes of order
$(l_P^{11})^2$.  However, at finite velocity the moduli space
approximation breaks down.  At probe momenta approaching $1/R_{11}$, the
probe is able to jump over the potential well defining the geometry
and move in a larger, non-singular space.

This enlarged space seems
to be a crucial element of sub-stringy ``D-geometry,''
as also seen in the promotion of space-time coordinates
to components of a matrix \WittenBound.   One might
conjecture that for every possible topology of space-time on small
scales, there is a specific linear sigma model which describes the
finite energy physics of each probe.  Clearly this would shed a new
light on such issues as topology change in string theory.  {}From the
work of Aspinwall, Greene and Morrison and of Witten
\refs{\AGM,\WitPhase}, we know topology change is possible
(and can be described by linear sigma models), but the D-brane
approach could give us a microscopic description.

It would be fascinating to extend the D-brane gauge theory
description to more non-trivial topology change, such as
the conifold transition \Strom.  It will also be very interesting to
see if D-brane moduli spaces on `non-geometric' compactifications
(such as Landau-Ginzburg) leads to a more geometric picture for them.

Despite the geometric nature of many of the results, we still find it
quite surprising to see that the metric is no longer a fundamental
degree of freedom at these short scales, but is a derived quantity.
Clearly much remains to be understood about this.

Some readers may have been bothered by what appears to be a preferred
role of flat space in our considerations: we reproduced non-trivial
metrics by solving D-brane gauge theories defined in flat space.  The
reason flat space appeared is that we chose to study D-branes moving
in a flat background space-time.  We could have taken a background
with curvature small in string units, and the world-volume theory
would have seen this metric instead.  This would lead to small
corrections in the problems we considered.

But this explanation raises more questions than it answers.
A prime question is whether there is some language unifying open and
closed string regimes.  Given that supergravity is not an appropriate
description in the sub-stringy regime, even foundational questions
such as the role of general covariance might have to be rethought.
Now the open string effective field theory description is also general
covariant -- one can always do field redefinitions to change the
coordinate system, and none is preferred, just as for sigma models of
fundamental strings \friedan.  But how should one think about the
larger configuration spaces in which these effective field theories
are embedded ?  

Many other questions remain to be addressed.  It is striking that the
ten-dimensional Planck scale $l_P^{10} \sim \gs^{1/4} l_s$ (far longer
than the length scales discussed in this paper) does not seem to
influence the dynamics analyzed here.  The connection of these ideas
to ten-dimensional gravitational physics, black holes, and information
will be important to explore.  The construction of D-brane systems
which are continuously connected to black hole solutions and the
computation of their Bekenstein-Hawking entropy \BlackHoles\ leads to
the exciting possibility that considerations such as these might play
a role in black hole physics~\toappear.

In our study of the $(2,1)$ bound state of 1-branes in type \IIb\ string
theory we found a characteristic size $\sim \gs^{1/2} \ls$ (up to logarithms).
This is essentially the 1-brane tension scale but its further
significance is as yet unclear.

Another set of questions concern 0-brane scattering behavior in the
relativistic regime, i.e., momenta $k >> 1/R_{11} \sim m_{0}$.  The
annulus calculation in \Bachas\ shows an inelastic ``stringy halo'' of
size $\sim l_s\sqrt{\log(kR_{11})} $ developing.  Is perturbation theory
accurate here?  Fixed-angle scattering is of magnitude $\exp(-1/\gs)$ in
this regime.  Do stringy nonperturbative effects, perhaps related to
0-brane pair creation come into play?  Do we have to consider the
transition from 0-brane quantum mechanics to 0-brane quantum field
theory in the relativistic regime?  What is the role of the hard scattering
caused by D-instantons \refs{\Green, \PolBdy} in this story?

More generally it will be important to develop a better 
eleven-dimensional M-theoretic understanding of the processes discussed here.
This will be important if we are to learn the lessons that this domain
of the theory is trying to teach us.

\bigskip
\centerline{\bf Acknowledgments}

It is a pleasure to acknowledge stimulating discussions with Tom
Banks, Mike Green, Jeff Harvey, Igor Klebanov, Juan Maldacena, Samir
Mathur, Joe Polchinski, Nati Seiberg, Andy Strominger and Lenny
Susskind.  This work was supported in part by DOE grant
DE-FG02-96ER40559, NSF grant PHY-9157016, and by a Canadian 1967
Science Fellowship.

\appendix{A}{Annulus amplitude and theta function identities}

We show that certain leading terms in the velocity expansion are
given exactly by the lightest modes in the open or closed string
channels.  The easiest way to check this is by series expansion of the
integrand; since it is a product of $\theta$ and $\eta$ functions it
will be a modular form of given weight under a congruence subgroup of
$SL(2,\BZ)$; the space of these is finite dimensional, so it suffices
to check it to a finite (low) order.

Let us check it analytically as well.
Some generalities:
$q=e^{-\pi t}$ (not to be confused with the q-brane dimension)
and $z=e^{2i\nu}$,
the theta functions are
 $\theta_{00}=\sum_{n\in\BZ} q^{n^2}z^n$,
 $\theta_{01}=\sum_{n\in\BZ} (-1)^n q^{n^2}z^n$ and
 $\theta_{10}=\sum_{n\in\BZ} q^{(n+1/2)^2}z^{(n+1/2)}$.
By using the heat equation
\eqn\heateqn{
{\p^2\theta(\nu|t)\over\p\nu^2} = 4i\pi{\p\theta(\nu|t)\over\p t}
}
the velocity expansion can be converted into an expansion in $t$
derivatives.  $\theta(\nu|t)$ and $\eta(t)$ have modular weight $1/2$,
and a $\nu$ derivative increases the modular weight by $1$.  The term
$v^k$ which could be a constant is the one with weight zero, a
function on the modular domain.  The weight will be zero when the
condition $k=4-(q-p)/2$ is satisfied, which was the case in which pure
massless closed string exchange could reproduce the lightest open
string result.  Thus the goal is to compute the modular function which
appears at order $v^k$.

Once we know that no tachyon divergence appears, we might be tempted
to argue that, since the integrand is a modular function which is
non-singular as $t\rightarrow\infty$ and $t\rightarrow 0$,
it must be a constant.
This idea might lead to a general proof, if one could argue
that singularities at other cusps were also ruled out.  We have not done
this and so will consider the possible cases explicitly.

We consider the product of \oploop\ and \opfermp\ or \opfermpneq\ as
a ratio of numerator over denominator.
Since the denominator will contain
$\eta(t)^k \sim q^{k/12}$ as $q\rightarrow 0$,
cancellation of tachyon divergences will
require the numerator to vanish in this limit as well, in other words
it will be a cusp form.  This is often strong enough to determine
it uniquely.

\subsec{Velocity expansion of \opfermp, $p=q$}

Let us first compute the leading term in the $q$-expansion,
due to the lightest open string states, at finite $v$.
We drop the kinematic term due to the bosons.
The rest is the $O(q)$ term in
\eqn\thetafour{\eqalign{
&\half\left[ \theta_{00}(0|t)^3 \theta_{00}(vt|t)
-\theta_{01}(0|t)^3 \theta_{01}(vt|t)
-\theta_{10}(0|t)^3 \theta_{10}(vt|t) \right]\cr
&= \half[(1+2q)^3(1+2q\cos \pi 2vt)
-(1-2q)^3(1-2q\cos \pi 2vt) \cr
&\qquad\qquad -16 q \cos \pi vt ]
 + O(q^2)\cr
&= 2q(3 + \cos \pi 2vt - 4\cos \pi vt) + \ldots \cr
&= 16 q \sin^4 {\pi vt\over 2} + \ldots \cr
&= v^4 q \pi^4 t^4 +O(v^6) .
}}

We now show that the $O(v^4)$ term is exact for all $t$.
The denominator, from \oploop\ and \opfermp, will be $\eta^{12}$,
a form of weight $6$.
The $v^2$ term is the first $\tau$ derivative of
$\theta_{00}(0|t)^4-\theta_{10}(0|t)^4-Q_1Q_2\theta_{01}(0|t)^4$.
For the brane-brane case $Q_1Q_2=1$, this is zero, while for
the other case this is $2\p\theta_{01}(0|t)^4/\p t$, a form of weight $4$.
The first non-zero term in the brane-brane case is
the $v^4$ term, a form of weight $6$.
Fourth powers $\theta^4$ will be forms of $\Gamma_0(4)$ and there is
a unique cusp form of this weight
(\koblitz, p. 146, problem 17).
In fact it is equal to $\eta^{12}$.

\subsec{Velocity expansion of \opfermpneq,  $q=p+4$.}

The contribution of the lightest open string states is the $q^{1/2}$
term in the fermionic partition function
\eqn\thetatwo{\eqalign{
\theta_{00}(0|t)^2\theta_{10}(0|t)^2
&\left[{\theta_{00}(v t|t)\over\theta_{00}(0|t)}
  -{\theta_{10}(v t|t)\over\theta_{10}(0|t)}\right] \cr
&= 4 q^{1/2} (1-\cos \pi vt) + \ldots\cr
&= 8 q^{1/2} \sin^2 {\pi v t\over 2} \cr
&= 2q^{1/2} \pi^2v^2 t^2 +O(v^4) .
}}
Now the $v^2$ term is exact.  $DD=8-q$, $NN=q-4$ and $DN=4$, so the
denominator is $\eta(t)^{(8-q)+(q-4)+2} = \eta(t)^6$.  The numerator is
$\theta_{01}(0|t)^{-2}\theta_{00}(0|t)^2\theta_{10}(0|t)^2
(\theta_{00}(v|t)/\theta_{00}(0|t)-\theta_{10}(v|t)/\theta_{10}(0|t))$,
and the potential cancels.  The first product of thetas is equal to
$4\eta(t)^6/\theta_{01}(0|t)^{4}$, leaving for the $v^2$ term
$4\pi{\p\over\p t} \log {\theta_{00}(0|t)\over\theta_{10}(0|t)} /
\theta_{01}(0|t)^{4} $, a ratio of weight $2$ forms, which is $-\pi/4$
(using an identity in \dkaz).

\appendix{B}{Non-parallel branes}

The discussion of section 2 can be extended
to the case of non-parallel branes; here
we outline the space-time interpretation for that case.

If the $p$ and $q$-branes 
are oriented orthogonally to the vector of minimum
separation, they can be treated by using T-duality
to relate them to parallel $p'$ and $q'$ branes.
Apply T-duality
to each world-volume dimension of the $p$-brane
which is not contained
in the world-volume of the $q$-brane, to `move it from 
one brane to the other'. 
That is, the dimensionality of the first brane decreases to $p'$ while
that of the other increases to $q'$.
This preserves $DD$, $NN$ and $DN$, and the result \oploop\ still applies.

The interpretation is a bit different.  On the open string side, the
modes are localized to the intersection region, of dimension $DD+1$.
For most purposes they can be thought of as living on a
sub-$p'$-brane of the $p$-brane (or `I-brane' \ghm).
On the closed string side, the field produced by the $q$-brane must be
integrated over the $p$-brane world-volume.  This integral converts
the $1/r^{q-7}$ falloff to a $1/r^{q'-7}$ falloff.

\bigskip
\hfuzz=20pt
\listrefs
\end